\documentclass[journal]{IEEEtran}

\usepackage[utf8]{inputenc}
\usepackage{cite}
\usepackage{setspace}
\usepackage{epsf}\epsfverbosetrue
\usepackage{graphics,epsfig}
\usepackage{graphicx,epsfig}
\usepackage{epsfig}
\usepackage{multirow}
\usepackage{alltt}
\usepackage{subfigure}

\usepackage{mathtools}

\usepackage{tcolorbox}
\usepackage{float}
\usepackage{url}
\usepackage{graphicx}
\usepackage{verbatim} 
\usepackage{footnote} 
\usepackage{latexsym} 
\usepackage{sidecap} 
\usepackage{wrapfig}
\usepackage{eso-pic}
\usepackage{fix-cm}
\usepackage{algorithm}
\usepackage{algpseudocode}
\usepackage{color}
\usepackage{wrapfig}
\usepackage{multicol}
\usepackage{flowchart}
\usetikzlibrary{shapes,arrows,matrix,decorations.pathreplacing,shapes.geometric,positioning,calc}

\newcommand{\foo}{\hspace{-2.3pt}$\bullet$ \hspace{5pt}}

\newcommand{\comst}[1]{\textcolor{black}{#1}}

\usepackage{layout}

\usepackage{amsmath}

\usepackage{textcomp}
\usepackage{multirow}
\usepackage{mathtools}
\usepackage{soul} 
\usepackage{amsmath} 
\usepackage{verbatim}
\usepackage{csvsimple} 
\usepackage{array}
\usepackage{subfig}
\usepackage[normalem]{ulem}
\usepackage{booktabs}
\newcolumntype{P}[1]{>{\RaggedRight\arraybackslash}p{#1}}

\usepackage{pifont}
\newcommand{\tick}{\ding{52}}%
\newcommand{\cross}{\ding{55}}%
\newcommand{\mult}{\ding{93}}%
\newcommand\tikzmark[1]{\tikz[remember picture] \node (#1) {};}

\newcommand{\newmod}[1]{\textcolor{black}{#1}}

\newcommand{\maaz}[1]{\textcolor{black}{#1}}
\newcommand{\mubashir}[1]{\textcolor{black}{#1}}
\newcommand{\rehmani}[1]{\textcolor{black}{#1}}
\newcommand{\cogcom}[1]{\textcolor{black}{#1}}

\begin{document}

\title{Differential Privacy in Cognitive Radio Networks: \mubashir{A Comprehensive Survey}}

\author{Muneeb Ul Hassan, Mubashir Husain Rehmani, Maaz Rehan, and Jinjun Chen
\thanks{M. Ul Hassan and J. Chen are with the Swinburne University of Technology, Hawthorn VIC 3122, Australia  (e-mail:  muneebmh1@gmail.com; jinjun.chen@gmail.com).}
\thanks{Corresponding Author: M.H. Rehmani is with the Munster Technological University (MTU), Ireland (e-mail: mshrehmani@gmail.com).}
\thanks{M. Rehan is with the COMSATS University Islamabad, Wah Campus, Pakistan (e-mail: maazrehan@gmail.com).}
}

\maketitle

\begin{abstract}

Background/Introduction: \cogcom{Integrating cognitive radio (CR) with traditional wireless networks} is helping solve the problem of spectrum scarcity in an efficient manner. The opportunistic and dynamic spectrum access features of CR provide the functionality to its unlicensed users to utilize the underutilized spectrum at the time of need because CR nodes can sense vacant bands of spectrum and can also access them to carry out communication. Various capabilities of CR nodes depend upon efficient and continuous reporting of data with each other and centralized base stations, which in turn can cause leakage in privacy. Experimental studies have shown that the privacy of CR users can be compromised easily during the cognition cycle, because they are knowingly or unknowingly sharing various personally identifiable information (PII), such as location, device ID, signal status, etc. In order to preserve this privacy leakage, various privacy preserving strategies have been developed by researchers, and according to us differential privacy is the most significant among them.

Methods: In this article, we provide a thorough survey on how differential privacy can play an active role in preserving privacy of cognitive radio networks (CRN). Firstly, we provide a thorough comparison of our work with other similar studies to show its novelty and contribution, and afterwards, we provide a thorough analysis from the perspective of various CR scenarios which can cause privacy leakage. After that, we carry out an in-depth assessment from the perspective of integration of differential privacy at different levels of CRN. Then, we discuss various parameters which should be considered while integrating differential privacy in CRN alongside providing a comprehensive discussion about all integrations of differential privacy carried out till date. Finally, we provide discussion about prospective applications, challenges, and future research directions.

Results/Conclusion: The discussion about integration of differential privacy in different CR scenarios indicates that differential privacy is one of the most viable mechanisms to preserve privacy of CRN in modern day scenarios. \cogcom{From the discussion in the article, it is evident that the proposed integration of differential privacy can pave the way for futuristic CRN in which CR users will be able to share information during the cognition cycle without the risk of losing their private information.}

\end{abstract}



\tikzstyle{decision} = [diamond, draw, fill=blue!50]
\tikzstyle{line} = [draw, -stealth, line width= 0.4mm]
\tikzstyle{elli}=[draw, ellipse, fill=red!30,minimum height=5mm, text width=5em, text centered]
\tikzstyle{block} = [draw, rectangle, fill=blue!15, rounded corners, minimum height= 10mm, minimum width = 20mm, text width=11em, text centered]

\tikzstyle{smallblock} = [draw, rectangle, fill=green!50, rounded corners, minimum height= 7mm, text width=5.5em, text centered]

\tikzstyle{midblock} = [draw, rectangle, fill=orange!50, rounded corners, minimum height= 7mm, text width=5.5em, text centered]

\tikzstyle{leftblock} = [draw, rectangle, fill=orange!50, rounded corners, minimum height= 7mm, text width=5.5em, text centered]

\tikzstyle{endblock} = [draw, rectangle, fill=yellow!50, rounded corners, minimum height= 7mm, text width=5.5em, text centered]

\tikzstyle{newblock} = [draw, rectangle, fill=yellow!50, rounded corners, minimum height= 7mm, text width=5.5em, text centered]


\section{Introduction}

The exponential surge in the usage of hand-held Internet of Things (IoT) devices \cogcom{caused a huge rise in wireless traffic.} Statista report revealed that the number of hand-held mobile devices is projected to reach up to 17.72 billion by the end of the year 2024~\cite{crnintro01}. This surge is causing an irregular usage of spectrum, which is further responsible to cause the issue of `artificial spectrum scarcity’~\cite{survey03}. Similarly, the worldwide analysis and measurement of spectrum utilization revealed that only \maaz{5-10 \%} of wireless spectrum is being used by licensed/authorized users~\cite{crnintro02}. \cogcom{All these factors lead researchers to investigate mechanisms which provide spectrum efficiency, and cognitive radio (CR) is one of them.} \maaz{Cognitive radio is a widely accepted model for efficient spectrum utilization~\cite{crnintro03}. CR was first coined by J. Mitola in 1999.  \cogcom{CR is an ambiance-aware intelligent wireless system}} which can dynamically adapt changes depending upon its surrounding RF environment~\cite{crnintro05}. \rehmani{CR works over the principle \maaz{of allowing CR users (also known as Secondary Users (SUs)), to access spectrum of licensed users (also known as Primary Users (PUs)), during idle time}.} This functionality of CR allows SUs to exploit underutilized bands of spectrum without causing any harmful inference to the communication of PUs~\cite{crnintro04}. Thus, SUs can dynamically access available spaces \maaz{in the spectrum band in order to manage it efficiently~\cite{crnintro08, newintroref01}}.\\
\maaz{To dynamically access the spectrum, SUs need} to follow complete cycle \cogcom{which involve spectrum sensing (SS),} spectrum analysis, and spectrum adaptation (also known as exploitation)~\cite{crnintro06}. SUs repeatedly carry out these functions in order to achieve the desired environmental conditions \rehmani{(A detailed explanation of functioning of CRN has been provided in Section.~\ref{Sec:Fundamental}). \maaz{All other functions, for example, spectrum auction (used to decide winner of spectrum allocation), etc, can be taken as a} sub-variant of the above-mentioned basic tasks.} These functionalities help SUs to find and select the best possible spectrum band in order to carry out seamless communication. \cogcom{Since these steps involve transmission of SUs data, these steps can be exploited by adversaries} to infer their personal data. \cogcom{For instance, multiple SUs are collected during collaborative spectrum sensing (CSS) by a fusion center (FC)} to get the best spectrum results. However, this FC can become an adversary and exploit private data of CR users~\cite{survey04}. Similarly, in case of spectrum auction, the centralized auctioneer can exploit the \cogcom{bidding privacy (BP) because it has all data} of multiple SUs and PUs from the bidding perspective. Therefore, it is important to protect the privacy of CR users by integrating some external privacy preservation mechanisms.\\

\begin{table*}[htbp]
\begin{center}
 \centering
 \footnotesize
 \captionsetup{labelsep=space}
 \captionsetup{justification=centering}
 \caption{\textsc{\\Comparison Summary of Related Survey Articles with their Contribution, and Various Scopes such as Differential Privacy (DP), Location Privacy (LP), Trading Privacy (TP), Spectrum Sensing Privacy (SSP), and Sources of Privacy Leakage (SoPL). Tick(\tick)~Shows that the mentioned topic is covered, Cross(\cross)~shows that the provided domain is not covered, and Asterisk(\mult)~shows that the particular topic is partially covered.}}
  \label{tab:surveytab01}
  \begin{tabular}{|P{2.5cm}|P{0.6cm}|P{0.7cm}|P{0.7cm}|P{6.5cm}|P{0.6cm}|P{0.6cm}|P{0.6cm}|P{0.6cm}|P{0.6cm}|}
  \hline

&  & & & & \multicolumn{4}{c}{\centering  \bfseries ~~~~~~~~~~~Scope} &\\
\cline{6-10}
\rule{0pt}{2ex}
\centering  \bfseries Major Domain  & \centering  \bfseries Ref. & \centering  \bfseries Year & \centering  \bfseries \maaz{Type} & \centering  \bfseries Contribution Summary & \centering \bfseries DP & \centering \bfseries LP & \centering \bfseries TP & \centering \bfseries SSP & \bfseries SoPL\\

\hline

\rule{0pt}{2ex}
\centering \textbf{Cognitive Radio Networks Vulnerabilities} & ~\cite{survey01} & 2013 & \maaz{Survey} & Comprehensive survey on security concerns and prospective threats linked to deployment of CRN & \centering \cross & \centering \mult & \centering \cross & \centering \cross & \mult  \\
\hline

\rule{0pt}{2ex}
\centering \textbf{Location Privacy in CRN} & ~\cite{survey02} & 2014 & \maaz{Book} & Provided an extensive survey on location privacy and its mitigation strategies in CRN. & \centering \tick & \centering \tick & \centering \cross & \centering \mult & \mult  \\
\hline

\rule{0pt}{2ex}
\centering \textbf{Security Threats \& Defences in CRN} & ~\cite{survey03} & 2015 & \maaz{Survey} & Carried out a detailed survey on threats \& countermeasures of secure for both primary \& secondary CRN users. & \centering \cross & \centering \mult & \centering \cross & \centering \mult & \mult \\
\hline

\rule{0pt}{2ex}
\centering \textbf{Location Privacy in CRN} & ~~\cite{survey04} & 2017 & \maaz{Survey} & A comprehensive survey on leakage sources and mitigation strategies for location privacy in CRN. & \centering \mult & \centering \tick & \centering \mult & \centering \mult & \tick  \\
\hline

\rule{0pt}{2ex}
\centering \textbf{Physical Layer Security} & ~~\cite{survey05} & 2019 & \maaz{Survey} & Extensively classified security techniques and applications for CRN physical layer. & \centering \cross & \centering \cross & \centering \cross & \centering \cross & \cross  \\
\hline

\rule{0pt}{2ex}
\centering \textbf{Physical Layer Security} & ~~\cite{survey06} & 2020 & \maaz{Survey} & A thorough discussion and analyzation of security attacks on physical layer of CRN is provided. &\centering \cross & \centering \mult & \centering \cross & \centering \cross & \cross  \\
\hline

\rule{0pt}{2ex}
\centering \textbf{Differential Privacy in CRN} & ~This Work & 2020 & \maaz{Survey} & A state-of-the-art survey on privacy leakage of CRN along with extensive evaluation from the perspective of integration of differential privacy in CRN. & \centering \tick & \centering \tick & \centering \tick & \centering \tick & \tick  \\
\hline

 \end{tabular}
  \end{center}
\end{table*}


\cogcom{In the quest of providing privacy in CRN, extensive research has been carried out by researchers to integrate different privacy preservation strategies with CRN.} For example, some works~\cite{survey01, survey02} proposed the use of anonymization techniques such as~\textit{k-anonymity} to preserve privacy of CR users. Similarly, some other works~\cite{ technique04, technique06, technique07} analysed the use of encryption to preserve privacy. Certain works~\cite{technique08, technique09, technique10, technique11} investigated the use of private information retrieval to protect private CR data. \cogcom{Alongside these techniques, some works also highlight the to use of obfuscation-based privacy (also known as differential privacy)} to preserve CR users’ privacy. Among all these mechanisms we believe that \cogcom{differential privacy is one of the most viable mechanisms to protect the privacy of CR users because of its dynamic and adaptive nature.} \\
The notion of differential privacy was first discussed by Cynthis Dwork in 2006 in order to protect privacy of statistical databases by adding random independent and identically distributed (i.i.d) noise in the data~\cite{crnintro07}. \cogcom{However, afterwards researchers working in the field of private CRN tried integrating this differential privacy notion with CRN at different aspects and they got fruitful results.} Since then, plenty of works highlighting the integration of differential privacy with CRN have been carried out in the literature. In this paper, we provide a thorough survey regarding integration of differential privacy at various scenarios of CRN in order to demonstrate the useful benefits that one can get via this integration. Similarly, we try to discuss various technical works that have already carried out this integration and published their work in the literature.

\subsection{Key Contributions of Our Survey Article}
\cogcom{Nevertheless, certain surveys from the perspective of security and location privacy of CRN have been presented in the literature,} but a specific survey that highlights the need, integration, functioning, and applications of differential privacy in CRN have not been presented yet. In this article, we carry out survey of state-of-the-art works involving the integration of differential privacy and CRN alongside providing certain use cases which can be beneficial for future researchers who are interested to explore this field further. To conclude, the key contributions of our article are as follows:
\begin{itemize}
\item We carry out comparative review of our survey article with previously published survey literature.
\item We provide an in-depth analysis over the scenarios in which privacy of CRN can be compromised. 
\item We provide a thorough analysis from the perspective of integration of differential privacy at different levels of CRN. Alongside this, we also in-depth survey of all state-of-the-art integrations involving differential privacy and CRN.
\item We provide a comprehensive analysis of parameters \maaz{which should be considered while incorporating differential privacy in CRN models.}
\item We highlight various challenges, open issues, and prospective future directions for researchers and scientists interested to explore the field of differentially private CRN.

\end{itemize}


\begin{table}[t]
\begin{center}
\small
 \centering
  \captionsetup{labelsep=space}
 \captionsetup{justification=centering}
 \caption{\textsc{\\ \footnotesize{List of Acronyms Used in the Article.}}}
  \label{tab:acrtable}
  \begin{tabular}{|P{1.5cm}|P{6.3cm}|}
  	\hline
  	\rule{0pt}{2ex}
	\textbf{Acronyms} &\textbf{Definitions}\\
  	\hline
  	\rule{0pt}{2ex} 
  	 BP & Bidding Privacy\\
  	\hline
  	\rule{0pt}{2ex} 
  	 CR & Cognitive Radio\\
  	\hline
  	\rule{0pt}{2ex} 
  	 CRN & Cognitive Radio Networks\\
  	\hline
  	\rule{0pt}{2ex} 
  	 CSS & Collaborative Spectrum Sensing\\
  	\hline
  	\rule{0pt}{2ex} 
  	 DP & Differential Privacy\\
  	\hline
  	\rule{0pt}{2ex} 
  	 FC & Fusion Center\\
  	\hline
  	\rule{0pt}{2ex} 
  	 IU & Incumbent User\\
  	\hline
  	\rule{0pt}{2ex} 
  	 IoT & Internet of Things\\
  	\hline
  	\rule{0pt}{2ex} 
  	 LP & Location Privacy\\
  	\hline
  	\rule{0pt}{2ex} 
  	 PU & Primary User\\
  	\hline
  	\rule{0pt}{2ex} 
  	 RF & Radio Frequency\\
  	\hline
  	\rule{0pt}{2ex} 
  	 SU & Secondary User\\
  	\hline
  	\rule{0pt}{2ex} 
  	 \cogcom{SVM} & \cogcom{Support Vector Machine}\\
  	\hline
  	\rule{0pt}{2ex} 
  	 \cogcom{SSP} & \cogcom{Spectrum Sensing Privacy}\\
  	\hline
  	\rule{0pt}{2ex} 
  	 SS & Spectrum Sensing\\
  	\hline
  	\rule{0pt}{2ex} 
  	 SoPL & Sources of Privacy Leakage\\
  	\hline
  	\rule{0pt}{2ex} 
  	 TP & Trading Privacy\\
  	\hline
  	\end{tabular}
  \end{center}
\end{table}


\subsection{Related Survey Works}
\cogcom{A comprehensive literature is available in the field of CRN, however, the aspect of privacy preservation in CRN is not much discussed} and only a very few surveys are available in this field. Similarly, this presented survey work is different from other surveys in a context \maaz{that it discusses} the integration, design requirements, functioning, and applications of differential privacy in CRN. To the best of our knowledge, \maaz{there is no prior work which covers multiple aspects of differential privacy in CRN}. None of the works discussed integration of differential privacy in cognitive cycles, therefore, \cogcom{we develop certain comparison matrices such as discussion about differential privacy, location privacy (LP), trading privacy (TP), SS privacy (SSP), and sources of privacy leakage (SoPL)}. \maaz{A comprehensive comparison of our survey with other surveys, based on the aforementioned parameters, is presented in Table~\ref{tab:surveytab01}.} \\
The first work discussing vulnerabilities of CRN from the perspective of security, privacy, and deployment threats have been presented by Bhattacharjee~\textit{et al.}~\cite{survey01}.  The authors first describe various architectural aspects of CRN \maaz{focusing weak links having security and privacy threats. Afterwards, they discuss} threats and vulnerabilities that CRN can face if they are attacked by some adversary. A brief book purely targeting location privacy of CRN have been presented by Wang and Zhang~\cite{survey02}. The work first highlights certain privacy preservation mechanisms, and then discuss the integration of \maaz{privacy mechanisms in CRN.} The major focus of the work is \cogcom{location privacy leakage during CSS and during database driven CRN.} Another comprehensive survey discussing the security threats and defences in CRN have been carried out by Sharma~\textit{et al.}~\cite{survey03}. The \maaz{work focuses over} highlighting security vulnerabilities in different layers of CRN. The article started \maaz{with discussion of CRN physical layer, then discusses security threats in upper and cross-layer CRN. Finally, authors provide in-depth insights on how game theory can play the role in enhancing the security of CRN}.\\
\maaz{Another survey by Grissa~\textit{et al.}~\cite{survey04} focuses location privacy leakage and its mitigation techniques in CRN.} \maaz{This work first discusses the sources of CRN which may cause privacy leakage, then presents CRN privacy preservation mechanisms and finally states CRN location privacy methods of spectrum discovery along with attack scenarios.} One more work providing an in-depth classification of security threats of physical layer of CRN is presented by Hamamreh ~\textit{et al.}~\cite{survey05}. \maaz{The article first classifies security techniques of CRN physical layer and then presents its detailed applications.} Another similar article discussing the basics, detection, functioning, and countermeasures of physical layer threats of CRN have been presented by Salahdine~\textit{et al.}~\cite{survey06}. The work first classifies all physical layer attacks in CRN, \maaz{then classifies and discusses attack detection techniques and the possible countermeasures.} \\
However, considering this discussion and after analysing all \maaz{possible surveys presenting privacy preservation in CRN, it can be concluded that the prevailing literature does not give an in-depth knowledge and analysis of differential privacy in CRN. Our presented work is the first one that covers  integration of differential privacy with CRN from an in-depth technical perspective.}

\subsection{Overview of the Article}

A brief list of acronyms used in our \maaz{survey article has been given in Table~\ref{tab:acrtable}. The rest of the} article is structured as follows: Section~\ref{Sec:Fundamental} provides a brief discussion about fundamental concepts and preliminaries of article from the perspective of differential privacy, CR and sources of privacy leakage. Afterwards, Section~\ref{Sec:Integration} provides an in-depth discussion about integration of differential privacy with CRN in different CR scenario. After that Section~\ref{Sec:Evaluation} talks about various performance matrices that can be used to evaluate integration of differentially private approaches in CRN scenarios. Then, Section~\ref{Sec:Technical} carries out an extensive survey of all technical works that have integrated differential privacy in CRN. Afterwards, Section~\ref{Sec:Applicability} provides in-depth discussion about applicability of CRN in various futuristic scenarios and applications. Similarly, Section~\ref{Sec:Challenge} provides insights about possible challenges and prospective future research directions. Finally, the article is concluded in Section~\ref{Sec:Conclusion}.

\begin{figure*}[]
\centering
\includegraphics[scale=0.6]{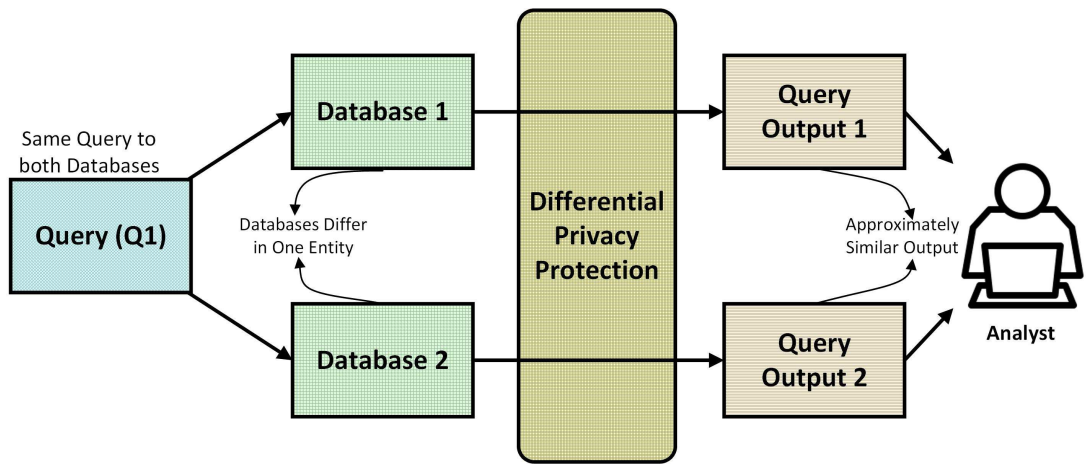}
\caption{\maaz{A Graphical Illustration of Functioning of Differential Privacy Mechanism in Two Adjacent Databases (adapted from~\cite{crnintro07}).}}
\label{Fig:DP}
\end{figure*}

\section{Fundamentals of Differential Privacy in Cognitive Radio Networks} \label{Sec:Fundamental}

\maaz{Since CRN serve as a viable solution to overcome spectrum scarcity issue, these have therefore been integrated with many domains, \cogcom{for example, smart grid, IoT, multimedia transmission,} transportation systems, healthcare and other~\cite{crnfund01}.} \cogcom{Although CR provides amazing features, it also has an issue of privacy leakage.} \maaz{Thus, CRN privacy protection} is also an important aspect that needs to be taken care of while implementing this paradigm. \maaz{In this section, fundamental concepts of our survey such as CRN, differential privacy, importance of privacy in CRN, privacy threats, and sources of privacy leakage in CRN is discussed. }

\subsection{Cognitive Radio and its Preliminaries}
\maaz{Discussion about is further categorized into five different aspects: (i) spectrum bands, (ii) CRN users, (iii) white space exploitation and utilization, (iv) CRN cognitive cycle, and (v) SS. }

\subsubsection{Licensed \& Unlicensed Bands of Spectrum}

Spectrum band, also known as frequency band, is used to carry out communication between devices, the band is widespread and ranges between 9 KHz to 3 THz depending upon the type of application~\cite{crnfund02}. The spectrum bands can be \maaz{further sub-divided into licensed spectrum band and unlicensed spectrum band~\cite{mubcrn01}}. The licensed band is allocated to licensed users who have paid licensing fees. These users can use the allocation frequency at any time without any inference and interruption. The licensed bands are usually allocated to telecom/Internet companies via auctions, who further allocate the bands to their customers~\cite{crnfund03}. Contrarily, unlicensed spectrum bands have been excluded from auction-based \maaz{sale/international licensing, therefore, these bands are used to carry out low-cost communication. These bands are} limited and their users are pretty large, therefore, these bands face the issue of heavy inference. \cogcom{Since, a large base of users compete for these unlicensed bands,} \maaz{National Authorities} regulating these bands carry out a conventional auction to manage these bands~\cite{crnfund02}. \maaz{In this way,} these authorities allocate a high paying user a fixed spectrum band and they can use it as a licensed user. \maaz{An analysis} indicates that the utilization of these bands \maaz{greatly} varies depending upon the geographical region. \maaz{For example, the overall utilization is 85\% at some places while it is 15\% at other places}~\cite{crnfund05}. This underutilization of spectrum leads to the formation of large unused white spaces, which are wasted. \maaz{To overcome} this spectrum wastage, researchers are getting benefit from the dynamic spectrum access functionality of CRN, through which CR users can access the unoccupied spectrum band and can leave at the time of PU activity.

\subsubsection{Primary \& Secondary Cognitive Radio Users}
\cogcom{The underutilization of spectrum can be controlled and reduced with the \maaz{help of CRN with very minimal level} of inference to licensed users. CRN mainly comprises two types of \maaz{users. First one is known as primary users (PUs), who are licensed and can access the spectrum at any time without any permission. The second} type of users are secondary users (SUs), \maaz{also known as CR users, who are unlicensed and can only access spectrum when it is unoccupied. SUs have to leave the spectrum in case PU arrives~\cite{crnfund06}. CR users continuously sense} the activity of PUs on the network and always look for white spaces. \maaz{Once an SU finds a white space, it moves to it and starts utilizing it for communication. In this way, SUs help to} overcome the issue of spectrum underutilization in an efficient manner. }

\subsubsection{White Space Utilization}
CR is an intelligent radio, which changes and \maaz{adapts to the changes according to environment. Similarly, extensive} research has been carried out in order to efficiently utilize the underutilized spectrum. Generally, \maaz{CRN can be further subdivided} into three paradigms on the basis of their spectrum utilization~\cite{crnfund07}. \maaz{First type of CRN is interweave CRN,} where CRN continuously senses the activity of PUs and access the network only when it is vacant. These are conventional CRN, which wait for the spectrum to become idle. Second type of CRN is underlay CRN, \maaz{which allows CR users to carry out operations alongside PU activity if the inference caused by CR users is less} than a specific threshold value. Third type of CRN is overlay, in which CR users overhear the continuous PUs transmissions and then use sophisticated algorithms of signal processing to enhance PUs performance, and in turn they get some additional patch of bandwidth which they can use to carry out their own transmission.

\subsubsection{Cognitive Cycle for White Space Exploitation}
\begin{figure*}[]
\centering

\includegraphics[scale=0.6]{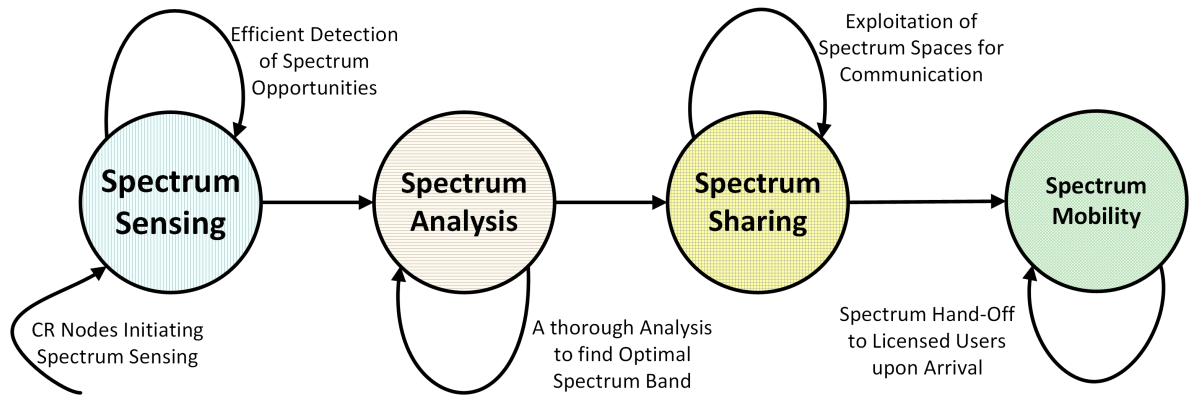}
\caption{\maaz{A Graphical Illustration of Functioning of Cognition Cycle for CR Nodes (adapted from~\cite{crnfund05}).}}
\label{Fig:CRN}
\end{figure*}

\cogcom{The spectrum utilization and exploitation phenomenon of CRN works in a stepwise manner, which is also known as the cognitive cycle. The cycle comprises four functional steps in which CR performs different actions to access spectrum in the most efficient manner.} The steps of the cycle are sensing, analysis, sharing, and mobility~\cite{crnfund08}. The details of these steps are given as follows:
\begin{itemize}
\item \textbf{Spectrum sensing} is the first step in the cognitive cycle. \maaz{Through SS,} CR nodes carry out efficient detection of spectrum opportunities around themselves in order to develop an initial idea of which spectrum band to access.
\item \textbf{Spectrum analysis} is the next step in which SUs select the best spectrum band from \maaz{the available bands.} \cogcom{The values collected via SS are used in spectrum analysis to choose a band in which inference to PUs is minimum and utilization is maximum.}
\item \textbf{Spectrum sharing} is the third step in the cognitive cycle which is used to exploit the chosen spectrum for communication. In this step, the decision of choosing the appropriate CR model (such as interweave, underlay, or overlay) is taken on the basis of data collected from previous two steps. At the end of this step, CR nodes can carry out transmission of their data via the selected spectrum band.
\item \textbf{Spectrum mobility} is the final step in the cognitive cycle and it involves the immediate mobility of SU nodes after detection of PUs. As the name suggests, this step is responsible for making the spectrum available to PUs by vacating the activity of SUs in case if \maaz{PU returns to resume} its activity.
\end{itemize}
\cogcom{Similarly, a figure to demonstrate functioning of the cognition cycle for CRN has been provided in Fig.~\ref{Fig:CRN}.}

\subsubsection{Spectrum Sensing and Access}
Since a \cogcom{large proportion of researchers focus on the integration of differential privacy during SS and access, let's demonstrate these steps in a bit detail for} better understanding. 
\paragraph{Collaborative Spectrum Sensing}
Usually, SS in CRN is carried out in a collaborative manner, in which all CR nodes collaborate with each other to generate the best SS outcome. This collaborative sensing is carried out to overcome the issue of shadowing \maaz{of PU. For instance, if a PU} is present on the rooftop of a high building and SU is on the ground level, then the presence or absence \maaz{of PU cannot be sensed with precision using a normal sensing mechanism due to low signal to noise ratio} (SNR) of the signal received from PU~\cite{crnfund09}. To overcome this issue of fading, multiple CR users collaborate with each other to carry out SS to ensure that they do not miss any specific PU. This whole process of collaborative sensing is done \maaz{with the help of a centralized} data collection \cogcom{center called FC. Generally speaking, FC collects data from all CRN and determines} the final values regarding presence or absence of PU in a particular region.

\paragraph{Database-Driven Spectrum Access}
According to database-driven spectrum access, data administrations are considered responsible for keeping up to date knowledge of spectrum and whitespaces in order to provide SUs with the most beneficial information~\cite{crnfund10}. A selected database collects necessary information from PUs including their usage times, tolerable inference, \maaz{available channels, allowed power transmission, and other useful information.} This information is then forwarded to SUs upon request, and the decision of joining or leaving a specific spectrum band is taken on the basis of this information. 

\subsection{Differential Privacy}

The perception of differential privacy as a medium to protect database privacy was first proposed by Cynthia Dwork in 2006~\cite{crnintro07}. This notion was later used by \maaz{researchers in almost every field to protect the privacy of their participants. For example, in auction, differential privacy has been used in auction to protect bid privacy. Similarly, it has been used in SS to protect location privacy.} To summarize, it will not be wrong to say that differential privacy is being applied to all real-life domains ranging from statistical databases to real-time decision analysis~\cite{crnfund11}. The formal definition of differential privacy form the perspective of two adjacent datasets $x$ and $x^\prime$ is as follows~\cite{crnfund12}:
\begin{equation}
\label{DPEqn01}
P_R[R(x) \in O_p] \leq \exp{\varepsilon} x P_R[R(x^\prime) \in O_p]
\end{equation}
In the above equation, $R$ is the randomized differentially private algorithm, $x$ and $x^\prime$ are two adjacent datasets, $P_R$ is the probability value for an outcome $O_p$ to be in range of function $Range(R)$. Furthermore, $\varepsilon$ is privacy parameter which is also known as privacy budget. The value of $\varepsilon$ is used to control the amount of noise which is going to be added in query result.\\
\maaz{Alongside $\varepsilon$, sensitivity is the other parameter} that plays an important role in determining noise value. \maaz{Sensitivity can be defined as the maximum difference an observer can get from the result of a query applied to two adjacent datasets} $x$ and $x^\prime$. The formal definition of sensitivity can be defined as follows~\cite{crnfund14}:
\begin{equation}
\label{DPEqn02}
\Delta S_q = \underset{x, x^prime}{max} ||f(x)-f(x^\prime)||
\end{equation}
Furthermore, various mechanisms of differential privacy have been proposed to calculate noise, and the two most famous among them are Laplace and Exponential. \maaz{These two use the pseudo random} noise generated from their respective database to perturb the query output. A detailed discussion of differential privacy from the perspective of mechanisms, composition theorem, sensitivity, and privacy budget can be found in~\cite{crnfund13}. Moreover, an illustrative explanation of differential privacy has been provided in Fig.~\ref{Fig:DP}.

\subsection{Importance of Privacy Protection in CRN}
\cogcom{Despite great advantages by CRN, they do suffer from a serious threat related to the privacy of its users. As discussed in the previous section,} CR nodes have to sense the spectrum in order to generate the environment map of PUs~\cite{crnfund16}. Similarly, queries and spectrum auctions are also carried out by FC to access and trade spectrum. However, while doing all these tasks, CR participating nodes have to report numerous \maaz{amounts of data to FC. Although FCs are usually trusted entities, in certain cases such as when there is an adversarial attack} on FC, the privacy of CR participants can be compromised~\cite{crnfund15}. For instance, the location of SUs and PUs can be compromised which can lead to serious consequences. \maaz{Similarly, the PUs and SUs usage and occupancy times can be analysed for malicious purposes.} Similar to that, bidding and asking prices can be analysed by adversaries for unethical actions (Detailed discussion on privacy sources and their countermeasures using differential privacy is given in Section.~\ref{Sec:Integration}).

\subsection{Adversary Models in CRN}

CR is a \maaz{diversified network, it therefore faces numerous types of adversarial attacks.} In this section, we categorize privacy related adversaries into four subtypes which cover approximately all types of adversarial attacks. The discussion about these adversaries is as follows:
\subsubsection{External Adversary}
\maaz{This type is one of the most prominent and dangerous as compared to others.} The adversary in this type includes any type of external intruder who is interested to get insights about the network in order to fulfil malicious objectives. For example, this adversary could be an external pharmaceutical company who is interested in finding out the number of CR users who visit hospitals/pharmacies frequently. They will do so by compromising the location of \maaz{PUs and SU. This company will collect all required information and can-do targeted advertisements. Usually, these adversaries operate in two ways, either compromised communication link or via compromised database.}
\paragraph{Compromised Communication Link} In this case, adversaries try to attack the communication link between SU/PU and central authority (usually FC). In this way, these adversaries try to overhear the communication between \maaz{SU and FC to infer their private information. One way is to launch man-in-the-middle attack~\cite{crnfund17} on the communication link, thus} adversaries can get required private information. Similarly, another attack in this type is exogenous attack~\cite{survey03}, in which an external adversary tries to jam the whole CR network to carry out malicious operations during this time.
\paragraph{Compromised Database} The second type of attack by \maaz{external adversaries could be in the form of compromising the FC database. In this attack, adversaries try to carry out a direct attack on FC database to get personal} information of participants including topology map of PUs and SUs~\cite{crnfund19}. This type of attack is usually carried out in database driven CRN. In collaborative sensing, this can also be done by intruding attack~\cite{survey03}, \maaz{where an intruder tries to get into the network externally by masquerading itself as a regular CR user, either for getting private information of CR nodes, or to inject falsified information in the network.}
\subsubsection{SUs acting as Adversary}
Alongside external adversaries, sometimes SUs can also act as adversaries and can play the role in compromising privacy of other CR participants. Similar to the intruding attack discussed in the above section, sometimes legitimate SUs can also \maaz{act as adversaries and try to leak into the privacy of other SUs}, PUs, and FC by collecting unnecessary data during cognitive cycle~\cite{survey04}. Nowadays there is a trend of decentralized sensing via blockchain, and all SUs try to reach a consensus in a decentralized manner~\cite{crnchallenge05}. This decentralized consensus is a great way to remove FCs, but it can also cause privacy issues because information reported by SUs is publicly visible to all other SUs, which can lead to harmful effects.

\subsubsection{PUs acting as Adversary}
\cogcom{Apart from SUs, sometimes, PUs can also act as adversaries. This rarely occurring case cannot be ignored especially in the case of} spectrum auction and trading. Since PUs are the authorities having excessive spectrum to sell, they also have objectives to enhance \maaz{their revenue. Some PUs therefore try to analyse the bids of SUs and try to take certain actions through which} they could increase their revenue in a tactical way~\cite{crnfund20}. 
\subsubsection{Service/Fusion Center Acting as Adversary}
The fourth type of adversary in our CRN modelling is centralized data centre-based adversary, which are also known as FCs. These FCs are usually trusted central entities and are designed to collect \cogcom{information from SUs during SS process to get the best spectrum. However, in some cases these FCs become adversaries. For example, FCs can sell the private data of their associated SUs and PUs to advertising organizations to earn extra profit. Similarly, these FCs can also analyse the data in a malicious way, which can even lead to the development of certain policies that may impact a certain group of participants.}

\noindent This discussion concludes that efficient privacy preserving mechanisms are required to protect CRN from these types of adversaries. In this perspective, differential privacy provides a strong privacy guarantee, which can be used to protect the privacy of CR users.

\subsection{Motivation of Using Differential Privacy in CRN}
Privacy preservation in wireless systems is a well-established field \maaz{and extensive research has been carried out in this regard~\cite{crnfund21}. In this section, a comparative analysis of these privacy preservation strategies with differential privacy. has been presented.}
\subsubsection{Encryption based Privacy Protection}
Since the advent of cryptography, encryption is being used to protect information in almost all fields of life~\cite{crnfund22}. Therefore, it will not be wrong to state that encryption-based privacy is the most traditional means to protect the privacy of any network/application. Encryption works over the phenomenon of key-based cryptography, in which a message is encrypted by a CR sender and then it is sent to CR receiver~\cite{crnfund23}. The receiver has the key which is used to decrypt the message. \maaz{The message in its encrypted form is known as ciphertext, which is unreadable and can only be decrypted by the person having the secret key. In this way, the message is protected from external intruders who cannot tap into the private communication of CR users.}\\
\maaz{Although, encryption provides strong privacy against external adversaries, but it is not an efficient privacy protection mechanism when internal participants become adversaries. According to our already discussed adversary models, encryption will only be helpful against external adversaries, but will not be much useful against the other three adversaries. There are certain encryption mechanisms which require strong computational power and a specific architecture for encryption}~\cite{crnfund25}. Such computational efficiency is hard to obtain in small CR nodes. Contrarily, differential privacy provides both these features, as it provides privacy protection from internal adversaries and is computationally less-expensive as compared to encryption.

\subsubsection{Oblivious Transfer based Privacy}
Oblivious transfer (OT) is also a popular means to protect privacy of CR nodes nowadays because it allows CR senders to send a message in multiple patches~\cite{survey04}. \maaz{In this mechanism, a message is broken down into multiple segments during transmission, which are then received and reassembled at the receiver as a single message. This approach is being used as a viable approach to carry out \cogcom{SS tasks in order to find out available spectrum without} compromising communication privacy. However, this mechanism suffers from similar issues of encryption-based privacy.}\\
Firstly, OT can only preserve privacy against external intruders, but cannot protect it from internal intruders. Secondly, the computational and communication overhead of this mechanism is quite high as compared to other mechanisms because multiple messages are transmitted at the same time, which incur high communication overhead due to collision, cohesion, and redundant rebroadcasts. \maaz{Differential privacy is however free from these issues, because the computational and communication overhead of differentially private messages is quite minimal,} and it also provides protection from internal intruders.

\begin{figure*}[]
\centering
\includegraphics[scale=0.6]{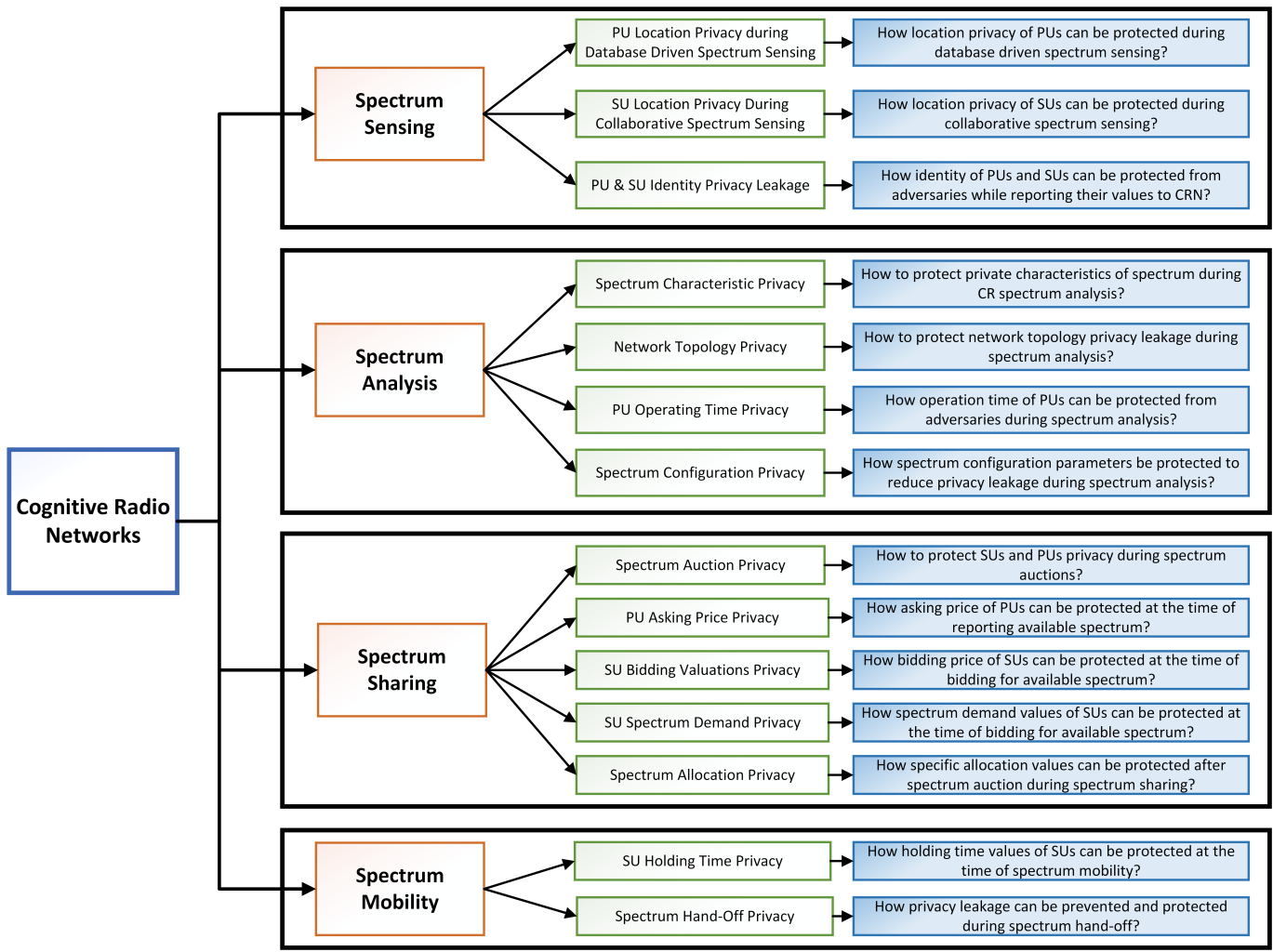}
\caption{\rehmani{Graphical Illustration of Sources of Privacy Leakage in Cognitive Cycle of Cognitive Radio Networks.}}
\label{Fig:Source}
\end{figure*}

\subsubsection{Data Anonymization-based Privacy}
Data anonymization is also a famous privacy preservation strategy which is being used to protect privacy during query evaluation of CR nodes~\cite{crnfund26}. In anonymization-based privacy, a dataset is anonymized by removing pseudo identifiable information from data before making it available for query evaluation. For instance, in CRN, FC can remove names and IDs of PUs before making the dataset available to SUs for database drive spectrum access. \\
This mechanism provides an effective solution for certain internal adversaries, but continuous experimentations have revealed that anonymized datasets can also be deanonymized by carrying out various linking attacks~\cite{crnfund27, crnfund27new}. Similarly, finding the best combination to remove from the dataset is also difficult, e.g., for some participants, coverage area could be confidential but not for others. Therefore, developing a consensus among participants on pseudo identifiable information is also tough. A major drawback of anonymization in CRN is that it requires a large database to operate, and in case if the database is not significantly big, then it can leak privacy. Contrary to all these aspects, differential privacy provides a dynamic mechanism which can be used to provide efficient privacy in these scenarios. Firstly, it is hard to identify the data protected via differential privacy due to its strong privacy guarantee. Secondly, differential privacy does not always require a large database, because pointwise differential privacy mechanism can also protect a single entity generated by CR nodes.

\begin{figure*}[t]
\centering
\captionsetup{justification=centering}
\begin{center}

\subfigure[]{
\includegraphics[width = 5.6cm, height = 3.1cm]{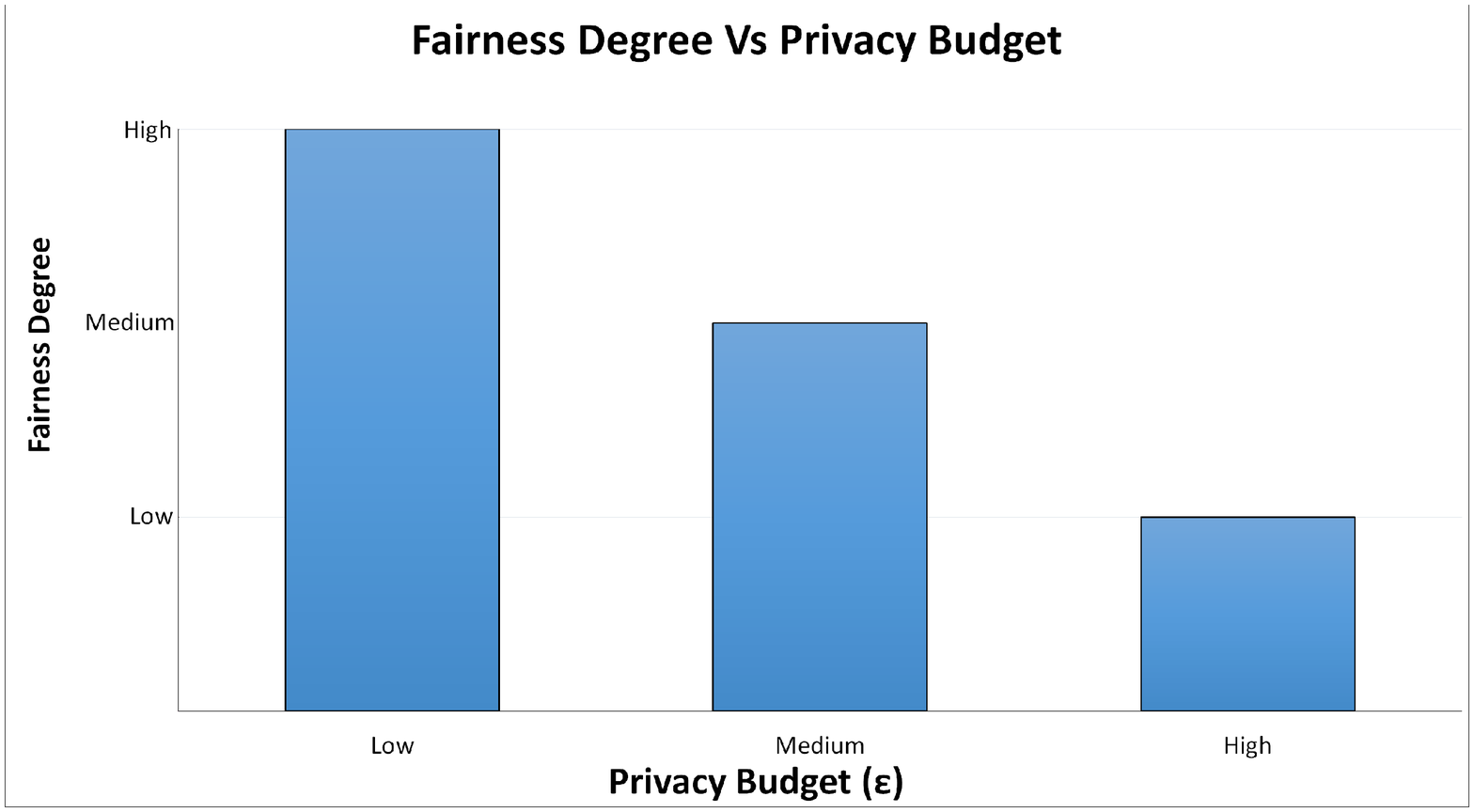}
}
\subfigure[]{
\includegraphics[width = 5.6cm, height = 3.1cm]{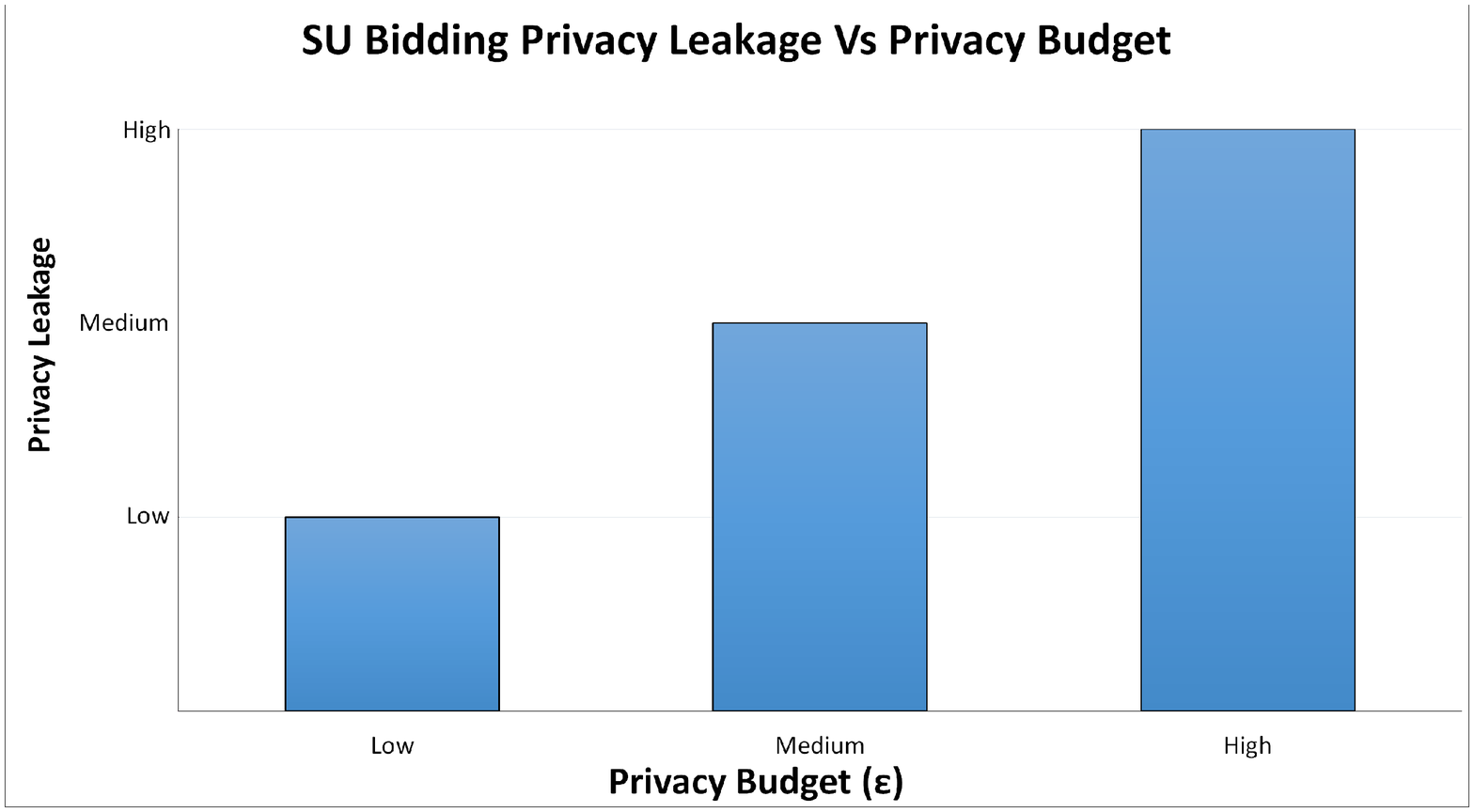}
}
\subfigure[]{
\includegraphics[width = 5.6cm, height = 3.1cm]{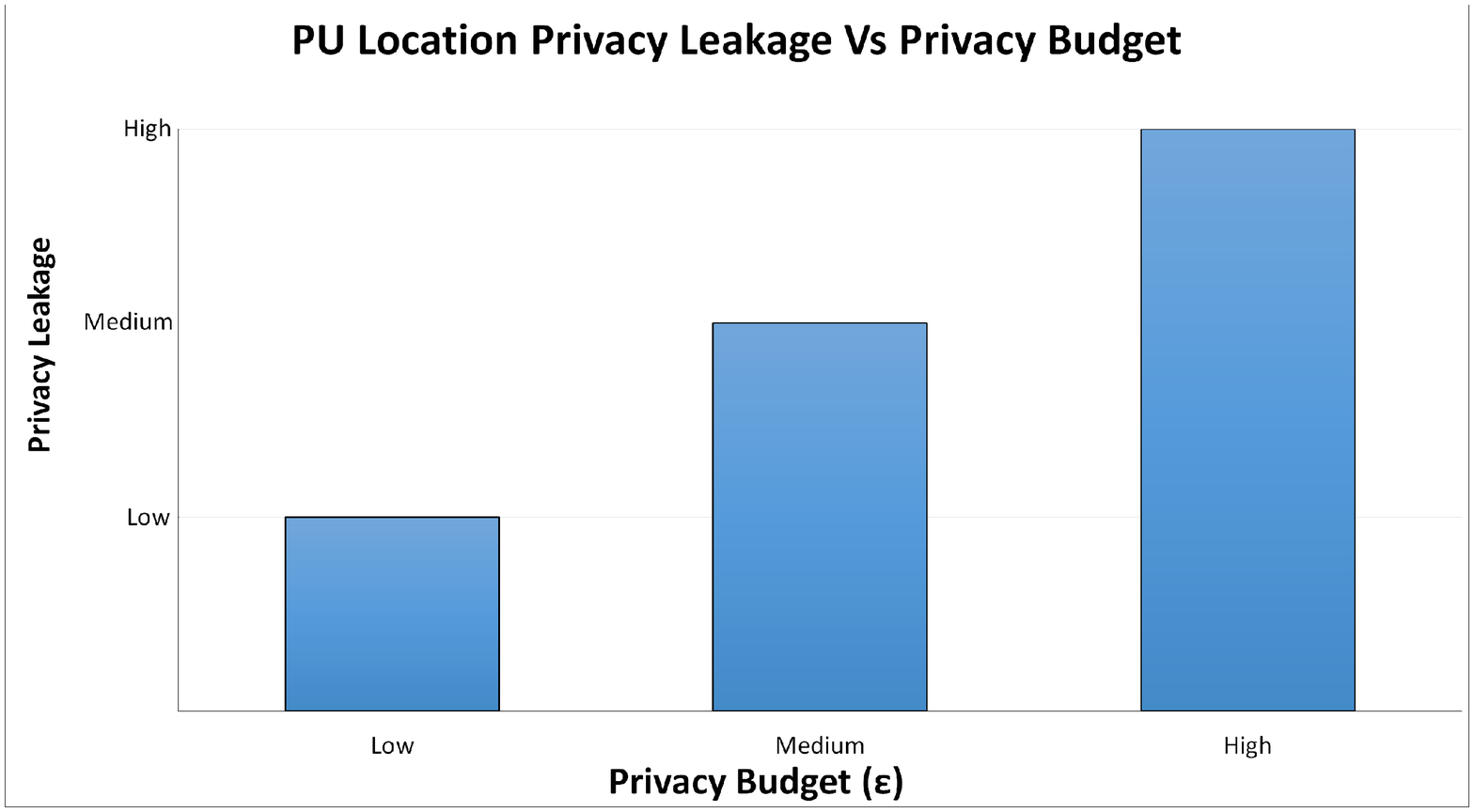}
}
\subfigure[]{
\includegraphics[width = 5.6cm, height = 3.1cm]{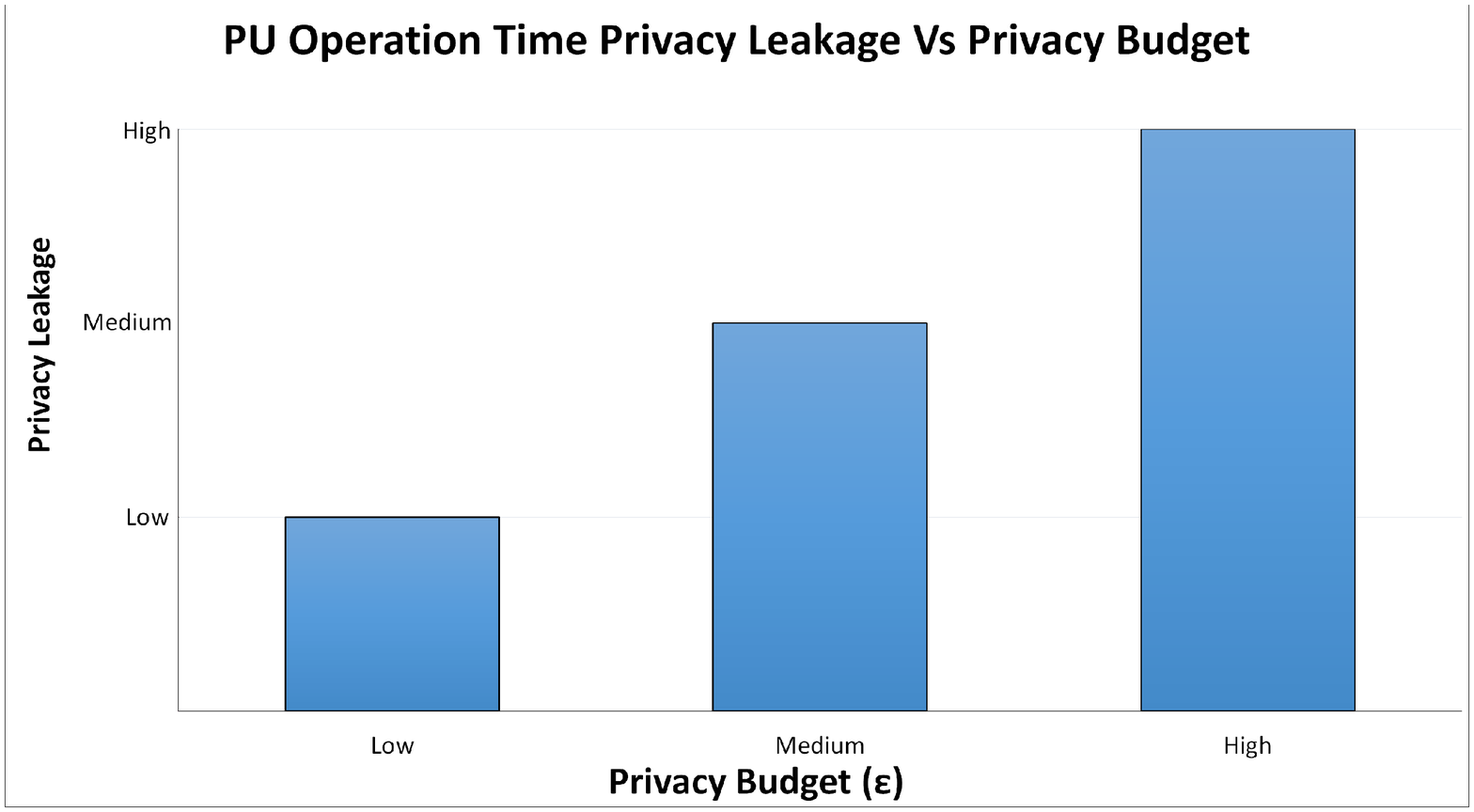}
}
\subfigure[]{
\includegraphics[width = 5.6cm, height = 3.1cm]{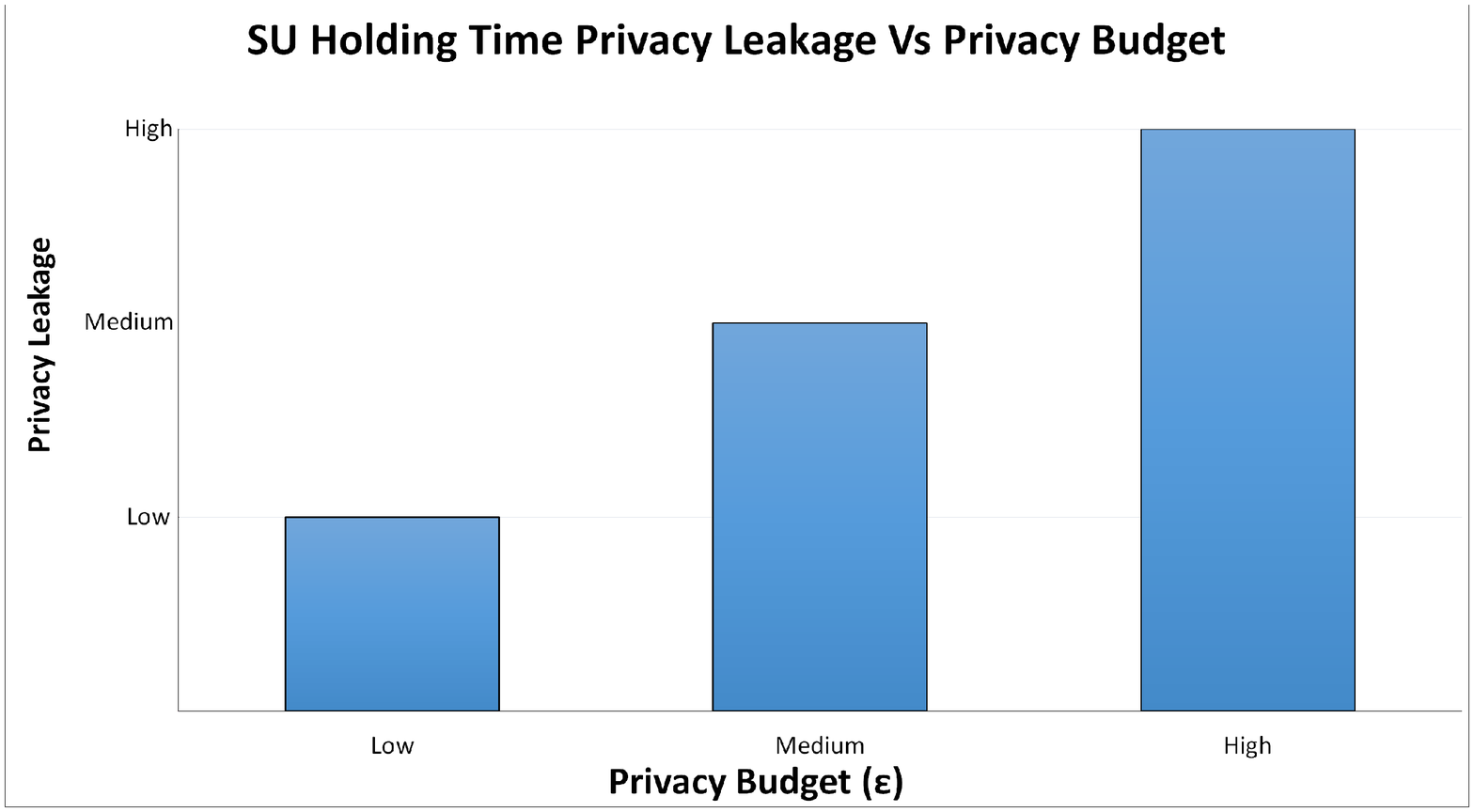}
}
\subfigure[]{
\includegraphics[width = 5.6cm, height = 3.1cm]{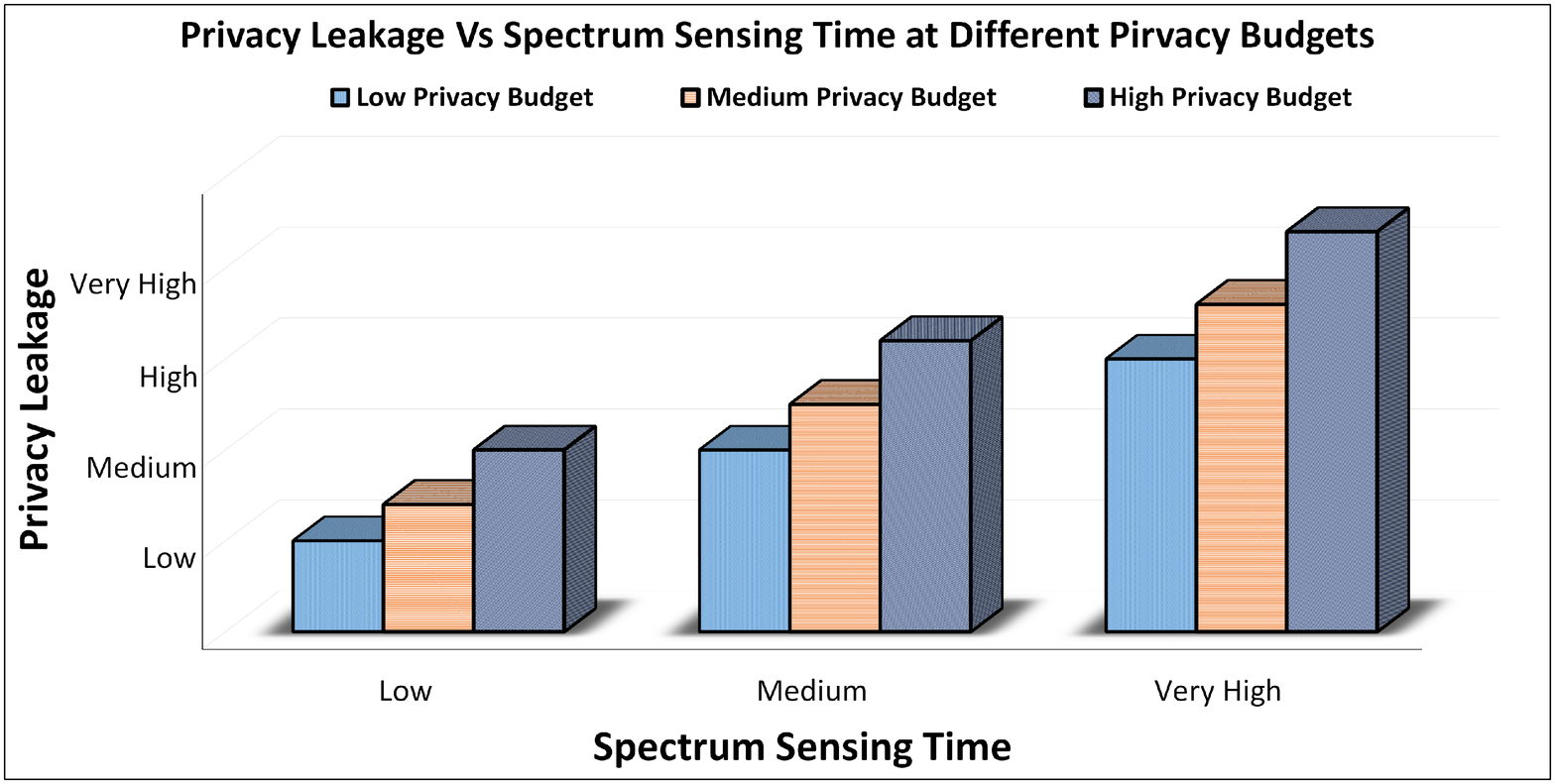}
}

\caption{\cogcom{Intuitive Analysis of Privacy Leakage in Various CR Scenarios at Different Privacy Budget ($\varepsilon$) Values.} \newline \cogcom{(a) Fairness Degree Vs Privacy Budget (adapted from~\cite{cogcomref12}) (b) SU Bidding Privacy Leakage Vs Privacy Budget (adapted from~\cite{bidpriv04}) (c) PU Location Privacy Leakage Vs Privacy Budget (adapted from~\cite{locpriv05}) (d) PU Operation Time Privacy Leakage Vs Privacy Budget (adapted from~\cite{locpriv04}) (e) SU Holding Time Privacy Leakage Vs Privacy Budget (adapted from~\cite{crnsource38})} \newline \cogcom{ (f) Privacy Leakage Vs Spectrum Sensing Time at Different Epsilon (adapted from~\cite{cogcomref13}).}}
\label{fig:IntuitiveGraph}
\end{center}
\end{figure*}

\section{\mubashir{Scenarios of Privacy Leakage} during cognitive cycle and Prospective Role of Differential Privacy} \label{Sec:Integration}

\maaz{In the previous section, a thorough discussion regarding the importance of privacy preservation in CRN from the perspective of SUs, PUs, and other involved participants is provided}. Moving further to classification, we discuss major sources of privacy leakage in CRN alongside in-depth analysis of how differential privacy can play its role in preserving privacy. \maaz{The sources are further categorized into four subtypes on the  basis of four major steps of cognitive cycle. An illustration of sources has also been provided in Fig.~\ref{Fig:Source}.} \cogcom{Moreover, in order to provide our readers, an intuitive overview of privacy leakage with respect to various prominent CR scenarios (such as fairness degree, SU bidding privacy, PU location privacy, PU operation time privacy, SU holding time privacy, and privacy in spectrum sensing time), we develop certain intuitive graphs, which have been presented in Fig.~\ref{fig:IntuitiveGraph}.}
\subsection{Privacy Leakage Scenarios during Spectrum Sensing}
\cogcom{SS is considered as a key functionality of CRN, as it provides information about available spectrum, which is the core for Dynamic Spectrum Access (DSA). SS in CR is performed by SUs in order to detect spectrum holes which they use to} carry out cognitive communication. \cogcom{The CRN environment is highly mobile and at certain times immediate decisions need to be taken in order to utilize the spectrum in the most efficient manner. Therefore, the aspects of accuracy, efficiency, timeliness, and decision making cannot be ignored~\cite{crnsource02}.} \maaz{In order to make quick decisions, a fine-grained data of participating nodes in CRN which provides efficiency on one hand, but on the other leaks the privacy of participants at multiple standpoints~\cite{crnsource03}.} \cogcom{In this section, possible scenarios that can leak privacy during SS of CRN are discussed.}
\subsubsection{PU Location Privacy from Databases-Driven Access}
According to \maaz{General Data Protection Regulation (GDPR),} location is a personal information and location of any participant/node cannot be traced without their approval~\cite{crnsource01}. \maaz{Similarly, Federal Communications Commission (FCC) has also} made it mandatory to preserve location privacy of PU nodes while designing CRN sensing techniques~\cite{crnsource04}. Database-driven CRN works over the phenomenon of query evaluation based spectrum access. It can easily be analysed that multiple queries from FC database can leak privacy of \cogcom{PUs who are involved in SS}. For instance, an adversary can try to launch multiple queries from database which collectively form a location inference attack. For example, if there is a single PU with name `X’ in a region `Y’, then the first query could be to find the number of PUs operating under a specific band (this will be done to find out the availability of spectrum). Similarly, the second query could be to find out the number of PUs in this specific region. Afterwards, the next \maaz{query could be, for example, to find out} the name of PU organization by saying `how many PUs have `X’ in their name’. \maaz{In this way, multiple queries can easily identify the presence and availability of a PU in a region.} This can lead to harmful consequences because other attacks can be launched to exploit such identified PUs. Therefore, location privacy of PUs during database-driven DSA should be protected. \\
\paragraph{Prospective Role of Differential Privacy} 
Since, database-driven DSA works over the phenomenon of query evaluation, so this location privacy issue of PU nodes can easily be protected via differential privacy. Since, differential privacy was designed to protect privacy of statistical databases~\cite{crnsource06}, it can be applied here to protect location privacy of PUs at the time of query evaluation. For instance, at the time of query from FC, a pseudo random noise generated from differential privacy distribution can be added into query output to ensure the randomness in the query output. However, the utility can also be maintained by controlling the privacy budget ($\varepsilon$) to a desired range~\cite{crnsource05}. Therefore, integration of differential privacy during query evaluation of database-driven DSA can effectively protect location privacy of PU nodes.

\subsubsection{SU Location Privacy During Collaborative Sensing}
\cogcom{SS is not only harmful for the privacy of PUs, but also poses a risk to the privacy of SUs} location~\cite{crnsource07}. \cogcom{In CSS, all SU nodes have to report the sensed values to centralized FC.} This collaboration results in an efficient and trustworthy sensing, but this also poses a high risk to the location privacy of SU nodes. For instance, untrusted, or compromised SUs can act as adversaries to locate the exact location of other SUs on the basis of the values received for SS~\cite{crnsource08}. For instance, location leakage of an SU can result in tracking of daily life activities of a particular SU. Therefore, considering the catastrophic outcomes of location privacy leakage, SUs are also concerned about their privacy protection. Considering this discussion, it can be concluded that privacy of SUs should be protected during CSS.\\
\textbf{Prospective Role of Differential Privacy:} Since SS in a collaborative manner is a critical aspect of CRN, it cannot be ignored. Therefore, protecting privacy during \cogcom{CSS is pretty important. Differential privacy is randomization technique} which can protect privacy from statistical databases to real-time reporting and it can be protected using pointwise differential privacy. Point-wise differential privacy means that any individual instance of data can be protected by adding pseudo random noise from differential privacy distribution~\cite{crnsource09}. Usually, Laplace and Geometric distributions are used to carry out point-wise perturbation of differential privacy. Similarly, if one adds a pseudo-random noise to protect location privacy of CR node at the time of SS, its privacy can easily be preserved. \maaz{Plenty of research is carried out to} integrate the phenomenon of differential privacy in location reporting~\cite{crnsource10}. The need here is to integrate this dynamic concept in the reporting aspect of SS in CRN.

\subsubsection{Identity Privacy Leakage During Crowdsourced Sensing}
Alongside location privacy leakage, another significantly important parameter is identity of PU and SUs. \cogcom{We briefly discussed this identity in query evaluation, but it is important to mention it separately as well, especially in the case of crowdsourced SS.} Since crowdsourced SS is a \maaz{combined activity, therefore, all CR nodes have to perform their best in order to get optimum results.} As all nodes share the information, this may lead to identity theft in case of an adversarial attack. For example, an adversarial node can pretend to be some other node if it gets to know all identity parameters of other nodes. In this way, many malicious acts can be carried out, ranging from adversarial decisions to DoS attacks. Therefore, it is important to protect identity privacy in SS alongside preserving location privacy.\\
\textbf{Prospective Role of Differential Privacy:} \maaz{In crowdsourced SS, the most important parameter to protect is identity of the sensing node, because this identity can further be backtracked to locate the specific node. However, at certain phases it is also important to check identity in order to ensure the integrity of the sensing report.} Differential privacy is a strong privacy guarantee that can also be used to protect privacy leakage in case of identity theft based attacks~\cite{crnsource12}. For instance, differential privacy can be combined with other provable encryption or proof mechanisms to provide a provable identity alongside preserving the actual values~\cite{crnsource13}. The similar concept can easily be applied for CR nodes during the aspect of SS. \maaz{Therefore, we believe adding calibrated differentially private noise with provable security mechanisms can be a viable solution to preserve identity theft privacy during crowdsourced SS.}

\subsection{Privacy Leakage Scenarios during Spectrum Analysis}
\maaz{Spectrum analysis comes after SS and is used to select an available spectrum}~\cite{crnsource14}. Similarly, the decision of bidding/trading for any specific spectrum is also taken on the basis of results from this step~\cite{crnsource15}. Spectrum analysis is broadly divided into two steps named as characterization, and reconfiguration. \maaz{First, the prospective scenarios in spectrum analysis which can become the source of privacy leakage are discussed and afterwards the role of differential privacy to overcome these leakages have been presented.}

\subsubsection{PU Privacy in Spectrum Characterization}
\maaz{After successful SS, SUs possess information about PUs and \cogcom{the available spectrum band which they can use to choose the most suitable. The scrutiny process is called spectrum characterization and it works as follows.} After collecting all data from the sensing step,} a list of spectrum bands is formulated involving various parameters such as path loss, RF environment, holding time, error rate, and switching delay~\cite{crnsource16}. These parameters are then used by SUs to determine the best available spectrum. \maaz{This process helps in the in-depth spectrum analysis but also leaks privacy. For example, the fine grained values, for example error rate and RF environment can easily be used to identify} the private characteristics of PUs.  Thus, privacy of PUs during characterization needs to be protected.\\
\textbf{Prospective Role of Differential Privacy:} Differential privacy is an advanced privacy protection mechanism which can be used to protect privacy during the characterization process. Firstly, differential privacy can be used to obfuscate the identity values so that the identity does not get leaked during characterization. Secondly, differential privacy obfuscation can further be applied to parametric values on the basis of requirement. E.g., a margin of error can be ignored in error rate or a margin of error in path loss can also be tolerated. So, the dynamic differential privacy algorithm can take advantage of this marginal error and can perturb data within this range in order to ensure privacy along with significant utility.

\subsubsection{Privacy Leakage in Learning Network Topology}
Similar to characteristics, network topology also carries certain private information in it. For instance, the geolocation of nodes can easily be inferred in case of unprotected network topology. In case of an adversarial attack on CRN, the adversary node may try to infer and collect all possible information collected via sensing, and after this inferring, the adversarial opponent may try to get much deeper insights about network topology~\cite{crnsource17}. This is done in order to find out the exact location of all PU nodes in the network. In case if the adversary successfully gets this personal information, then it can launch all attacks associated with location privacy, as discussed above. Therefore, it is important to protect privacy of network topology before publishing data during spectrum analysis.\\
\textbf{Prospective Role of Differential Privacy:} Since network topology depends upon multiple aspects, such as geo-location, \maaz{signal strength, they are combined together to form a complete topology. To protect} privacy during network topology formation, one therefore needs to protect individual parameters. The obfuscation of differential privacy can efficiently protect this by adding pseudo random point-wise noise to all parameters individually while considering the overall network utility~\cite{crnsource18}.

\subsubsection{PU Operation Time Privacy}
During spectrum characterization, an aspect of operation time cannot be ignored from a privacy viewpoint. Operation time is the total activity time of primary user, which is also known as primary user activity time~\cite{crnsource19}. Usually, during development of CR mechanisms researchers use various PU activity models to analyse the mechanism behaviour~\cite{crnsource20}. These PU models are used to determine the presence, absence, and functioning of PU nodes. However, in real-life scenarios, if one has this much fine-grained information about activity and presence of any node, then it can cause serious consequences to its privacy. For instance, if an adversary has \maaz{fine-grained information about PU activity, it can easily infer a daily schedule. For example, if an adversary is active then it means it is at a particular place, and if the spectrum band is unused, then the licensed user must be sleeping} or doing other tasks. This data is fed into machine/deep learning models, which further train themselves and try to predict the accurate lifestyle. Therefore, privacy of PU operation time needs to be protected before sharing this information to SU nodes.\\
\textbf{Prospective Role of Differential Privacy:} Protecting real-time lifestyle privacy is one of the key roles of differential privacy. Due to this advantage, differential privacy is being used by researchers in multiple aspects, for example, protecting privacy real-time smart metering data, protecting privacy of real-time EVs data~\cite{crnfund13}. Similarly, this aspect can also be applied to protect PU operation time privacy. For example, PU operation time values can be perturbed using differentially private noise from the distribution, and this noise can be calibrated according to privacy budget and data sensitivity. 

\subsubsection{Spectrum Reconfiguration Parameter Privacy}
After successful categorization of desired channel, the step of spectrum reconfiguration arises in which parameters of the transceiver of SU are configured according to the given condition~\cite{crnsource21}. This involves configuration of power, bandwidth, frequency, and other communication technologies. \cogcom{Although these are important parameters, which need to be configured properly, they in return can also leak privacy. For example, power control parameters can be used to figure out signal to interference ratio (SINR),} which can further be used to geo-locate the particular SU. Therefore, it is important to protect privacy of these parameters as well at the time of system reconfiguration.\\
\textbf{Prospective Role of Differential Privacy:} Differential privacy is a viable solution to protect the configuration data. Since, certain configuration parameters, for example frequency and values, are pretty strict and cannot be changed, therefore differential privacy can be combined with some provable mechanism to ensure privacy~\cite{crnsource12}. Furthermore, in case of parameters which can bear some noise of error, point-wise differential privacy can be integrated to protect their privacy. 

\subsection{Privacy Leakage Scenarios during Spectrum Sharing}
Spectrum sharing/decision is the third step after spectrum analysis, this step is further divided into three major steps involving allocation of resources, accessing of spectrum, and trading of spectrum~\cite{crnsource24}. To demonstrate it further, after spectrum analysis, CR nodes have the choice to choose the best available spectrum, and in order to do so, they first participate in spectrum trading to win the best available spectrum slot. Afterwards, the desired resource is allocated to them by FC or by some other server, and after completion of these two steps, they can access the spectrum for communication. These steps involve plenty of informational parameter exchange, which can lead to CR users. In this section, various cases of privacy leakage in scenarios involving spectrum sharing have been discussed.
\subsubsection{Privacy Leakage in Spectrum Auctions}
\maaz{Spectrum auction is a whole new world involving mathematical models.} For instance, game theory has been applied to it to achieve better results during auctions~\cite{crnsource25}. Similarly, for the majority of auctions, equilibrium is usually evaluated to get best results~\cite{crnsource26}. Alongside this, machine learning is also being applied to spectrum auctions to predict best outcomes~\cite{crnsource27}. Since this step has been explored a lot, the risk of privacy leakage has also increased a lot, and plenty of attacks on auction mechanisms have been developed in the past. \cogcom{In this section, we also discuss certain sub aspects of auctions such as BP.}\\
\textbf{Prospective Role of Differential Privacy:} \maaz{Since spectrum auctions are vulnerable to privacy attacks, therefore, privacy preservation is required during this process. To preserve auction privacy, differential privacy obfuscation is a viable solution because with differentially private auctions, one can still enhance social welfare of auction alongside preserving differential privacy. Similarly, in case of spectrum auctions, differential privacy can be used to protect privacy of both SUs and PUs acting as spectrum sellers and buyers, respectively.} (A detailed discussion about these mechanisms have been given in Section. ~\ref{Sec:Technical}).

\subsubsection{PU Asking Price Privacy}
Submitting available spectrum slots alongside asking price is one of the first step of the auction process. In this step PU node submits the available spectrum value alongside its asking price to FC or centralized server, which is further displayed to CR nodes to collect bids~\cite{crnsource28}. However, these values are critical, as they contain \maaz{information about PU spectrum usage, and an adversary can get insights that when a PU is vacant, or occupied.} By this analysis, the adversary can plan an adversarial activity or can launch an attack on the basis of previous knowledge. Similarly, an adversary can get insights about financial condition/dependencies of PU on the basis of total spectrum band and the price he asked for it. Therefore, it is important to protect asking price privacy before publicizing these values.\\
\textbf{Prospective Role of Differential Privacy:} Exponential shuffling, a mechanism of differential privacy can be used to introduce randomness in asking price string~\cite{crnsource22}. \cogcom{Similarly, the Laplace mechanism of differential privacy can be used to protect the privacy of asking price while managing social welfare of the complete auction process.} Similarly, other ways could be to integrate differential privacy with some provable mechanism to protect identity indirectly alongside preserving privacy of PU nodes.

\subsubsection{SU Bidding Privacy}
Apart from PUs asking price privacy, it is important to protect valuations/bids of buyers during the auction process. Valuations for a specific spectrum band is personal information and the majority of CR buyers do not want their private valuations to get leaked~\cite{crnsource29}. To overcome this, researchers integrated sealed bid auctions in CRN, but modern machine learning based attacks have caused privacy leakage even in sealed bid auctions. Therefore, it is important to protect privacy of bidding price alongside preserving privacy of asking price.\\
\textbf{Prospective Role of Differential Privacy:} Private valuation of auction process can easily be protected using randomized differential privacy mechanism~\cite{crnsource30}. For SUs, differential privacy works from the perspective of addition of random noise in the set valued data, which plays an important role in hiding the actual valuation. For instance, a  randomized valuation within the range of social welfare maximization can be generated for auction with the help of differential privacy. This randomized valuation ensures that the privacy does not get leaked and the social welfare for auction still remains positive. 
\subsubsection{SU Demand Privacy}
Alongside valuations, the amount required by a particular SU is also personal, because no CR node wants to reveal the exact spectrum usage to any adversary~\cite{crnsource31}. \cogcom{This is because these spectrum usage values can further be used to carry out burglaries or other attacks at non-usage/idle times. Similarly, the demand of a particular SU does also signify the financial situation and other similar aspects. Therefore, SUs usually try not to reveal their actual demand to auction places where there is a chance of adversarial attacks.} Considering this discussion, it can be claimed that if the demand of SUs can be protected, a large gain in spectrum trading can be seen because SUs will be able to participate in auction without the risk of losing their private information.\\
\textbf{Prospective Role of Differential Privacy:} Demand by a particular SU is a personalized information, which it needs to be protected~\cite{crnsource32}. This value can easily be protected by dynamic differential privacy mechanisms. Differential privacy can be integrated at multiple steps during this demand protection process. \maaz{The most significant way to overcome this issue is to integrate differential privacy with some provable mechanism. By doing this, one will be able to show perturbed demand alongside having a protected demand at the backend for verification and allocation.} \cogcom{Another way could be to collect variable demand from SUs and then find out the finalized value via an obfuscation mechanism. Third way could be to use the Exponential obfuscation mechanism in a way that does not harm the utility but introduces a selection randomness. All these mechanisms can be used to protect demand privacy in a viable manner.}

\subsubsection{Privacy Leakage in Spectrum Allocation}
After careful collection of asking price and bids, the next step is the allocation of spectrum. This step is usually carried out via some game-theoretic auction model. Some prominent auctions used in CRN include double auction, VCG auction, Dutch auction~\cite{crnsource33}. These auction models analyse all collected values and find the auction winner according to the prescribed process. \maaz{Although, highest} bidder wins the best spectrum in these auctions, but this can cause privacy leakage at various levels. For instance, the identity privacy of the winner can be leaked by analysing the auction result~\cite{crnsource34}. Similarly, the identity privacy of sellers gets leaked and it can be analysed that which seller has a specified amount of available spectrum. Alongside this, the financial situation of PU and SU nodes can be analysed by adversaries as a result of this spectrum. Similarly, the number of available channels involved in auctions can also be leaked during allocation. Therefore, designing a privacy preserving spectrum allocation model should be considered while developing CR auctions.\\
\textbf{Prospective Role of Differential Privacy:} Significant research on integration of differential privacy in auction allocation have been carried out and these research works have showed that differential privacy protection is a suitable method to protect privacy during auctions allocation~\cite{bidpriv07}. From CRN perspective, there is a need to design certain works which ensure truthful and private allocation alongside enhancing social welfare. In this way, the Laplace mechanism of differential privacy can play a significant role to ensure identity and multi-channel privacy at the time of allocation by integrating controlled randomized obfuscation.

\subsection{Privacy Leakage Scenarios during Spectrum Mobility}\label{SubSec:Mobility}
The fourth and final step in the cognitive cycle revolves around movement of \maaz{SU nodes at the time of PU arrival called as spectrum mobility~\cite{crnsource36}. In this step, first of all, PU tries to access the licensed spectrum back, which in return forces SU nodes to leave the spectrum immediately. Then SU vacates the spectrum and stops communication on it. SUs then looks for another available spectrum or waits for the PU to stop the communication again in order to resume their communicationn~\cite{crnsource37}. In both cases, privacy gets leaked at multiple events ranging from request to hand-off. Let's discuss} that how differential privacy can play its role to mitigate these privacy risks.
\subsubsection{SU holding time Privacy}
When an SU occupies a spectrum, it holds the spectrum until the PU arrives, or it holds the spectrum until the required communication need is fulfilled~\cite{crnsource38}. In both cases, SU does not want other adversaries to know the exact amount of time the spectrum was held by it. For instance, if the specific time gets leaked, then the adversary can find out that the particular SU held the spectrum on a particular place for an `X’ amount of time which can further lead to harmful events. Therefore, protecting privacy of SUs holding time is important and specific privacy preserving mechanisms should be designed for this purpose.\\
\textbf{Prospective Role of Differential Privacy:} \maaz{Channel holding time can be used to infer private information of SU, it should therefore be} protected by designing privacy preserving mechanisms. In this case, differential privacy can play an active role by integrating Laplace based obfuscation in values of holding time~\cite{crnsource39}. This obfuscation will ensure that adversaries will not be able to predict with confidence about the presence or absence of a specific SU on a spectrum. Similarly, at the time of sharing holding time, the exponential randomness can be used to ensure randomized response in order to protect privacy of SU.

\subsubsection{Privacy Leakage during Spectrum Hand-Off }
\maaz{When PU arrives, SUs are forced to leave the spectrum causing privacy leakage at multiple levels.} Firstly, PU may know that a particular SU is within its specified region. Secondly, SUs might have to carry out SS again to find the new appropriate spectrum, which again opens the door for all privacy leakage scenarios during SS. \maaz{Thirdly, alongside SU, the identity privacy of PU is also at risk, because SU can infer the presence of PU in its  region by visualizing the request~\cite{survey04}.} Therefore, it is important to integrate a privacy preserving mechanism at the time of spectrum hand-off.\\
\textbf{Prospective Role of Differential Privacy:} Spectrum hand-off comprises of multiple steps, so a simple obfuscation mechanism will not be enough to protect privacy as it will only cover a single aspect. Therefore, there is a need to design such differential privacy mechanisms which protect the privacy of multiple parallel events to ensure a trustworthy CRN. For instance, a Laplace based location privacy mechanism can be used to protect the privacy of SUs location~\cite{crnsource39}. In this mechanism, differential privacy can be used to introduce randomness in the location reporting to PU. Secondly, the sensing request should be done in a randomized way to protect this request privacy~\cite{crnsource40}. Afterwards, from the perspective of PU, an Exponential mechanism can be used to show randomised PU to SU instead of accurate identity. Similarly, the location privacy of PU at hand-off can also be protected by integrating Laplace obfuscation at the time of hand-off request.

\subsection{Summary}
In this section, we analyzed privacy leakage scenarios during cognitive cycle which can be exploited by adversaries to infer privacy of CR participating users. We analyzed all four steps involved in cognitive cycle ranging from sensing, and analysis to sharing and mobility. In SS, majority of sources are linked with location privacy. For instance, location privacy of SUs and PUs during collection of their values. \cogcom{Contrary to this, the sources in spectrum analysis step are more related identical threats such as characteristics privacy and network topology privacy.} Moving further to spectrum sharing, it can be visualized that literature is more titled towards privacy leakage during different steps of trading. Finally, in spectrum mobility, the privacy leakage is discussed from the perspective of hand-off, which involves both identity and location privacy. \\
Moving further to the mitigation of these privacy issues, we also provide an in-depth discussion about each possible scenario that how differential privacy can play its role in overcoming privacy issues at different steps. We have provided a detailed discussion from the perspective of integration of both Laplace and Exponential mechanisms of differential privacy in CRN. From the discussion it can be deduced that differential privacy is a dynamic privacy preservation strategy that can play a significant role in all four cognitive cycle steps. And it can be used by researchers in multiple CR scenarios, where there is a risk of privacy leakage.

\section{Performance Matrices for Evaluating Differentially Private CRN Mechanisms} \label{Sec:Evaluation}

\maaz{In the previous sections, we highlight significance of differential privacy in preserving privacy of CRN. This section highlights performance parameters that should be taken care of while developing of differentially private CR mechanisms. \cogcom{This section divides performance matrices into three categories, first are the matrices which are required while designing differential privacy techniques; second are those matrices which should be taken care of to avoid attacks; and third are those matrices which should be focused in CR applications.}}
\subsection{From the Perspective of Privacy Preservation}
\maaz{There are three important parameters which needs to be considered while developing of differentially private CRN works. These are discussed below.}

\subsubsection{Degree of Privacy}
While designing any privacy preservation mechanism it is important to figure out the degree of privacy required by application. For example, some applications might need a high level of privacy, but they can compromise on utility, for example, EV battery status reporting~\cite{crnmatric16}. Contrarily, some applications might need a high level of utility but can compromise a little bit on privacy, for example, industrial manufacturing. \cogcom{Similar is the case in CRN, where some aspects in the cognitive cycle might require a higher level of privacy. For example, location reporting during SS might need a high level of privacy due to involved location inference risks. However, some aspects might require a high level of utility,} such as available frequency values from a particular PU. Considering this discussion, it is important that one should determine the degree of privacy before designing and evaluating the privacy preserving model for CRN. \maaz{Due to differential privacy noise addition, the degree of privacy can be further divided into: (a) degree of privacy due to noise, and (b) degree of privacy due to anonymity.} The detailed discussion on these parameters is given below.

\paragraph{Degree of Privacy due to Noise}
While designing a differential privacy mechanism, it is important to figure out the level of noise which is required for the specific cycle. In differential privacy, two factors actively contribute to determine noise, first one is privacy budget ($\varepsilon$), while second is sensitivity ($\delta$). In a differentially private CR mechanism, $\varepsilon$ is used to determine the value of noise which is going to add in the output result. This $\varepsilon$ is chosen after careful consideration and is usually backed by a strong theoretical guarantee. For instance, specific theoretical contributions have been carried out to choose the appropriate $\varepsilon$ value~\cite{crnmatric09}. This $\varepsilon$ is inversely proportional to noise level, which means the higher value of $\varepsilon$ provides less privacy and low values of $\varepsilon$ provides high privacy values. Similarly, in different CR scenarios different $\varepsilon$ values are required, which is determined at the time of mechanism design. For instance, in case of location reporting during SS, the desired $\varepsilon $ value is pretty low, which ensures that the chances of location value getting leaked are pretty less. \maaz{On the other hand,} if one is determining final price of spectrum band during auction, only a minor level of noise can play the required role. Therefore, in such cases even the high value of $\varepsilon$ can fulfil the requirement. \\
Similarly, the sensitivity $\Delta$ value of a differentially private mechanism is usually determined on the basis of data. Formally, sensitivity is defined as a maximum possible impact of one record in accordance with all neighbouring datasets. This value is used in the noise addition and plays a significant role in determining the noise value. For instance, a CR database in which the participants have such values which differ a lot with each other might have high sensitivity value. \maaz{On the other hand,} a CR database in which the difference is pretty small, will have low sensitivity. Therefore, it is important to determine the sensitivity of model before implementing it. Certain works also highlighted dynamic varying sensitivity on the basis of dynamic data, but even in that case a predetermined method to choose appropriate sensitivity is required. 
\paragraph{Degee of Privacy due to Anonymity}
Another parameter similar to noise level is anonymity level, which basically is the level of privacy/anonymity after noise addition. This level is determined by comparing the anonymized dataset with original dataset. Similarly, certain works also mentioned it as degree of privacy leakage. \cogcom{In differentially private CRN, it can be termed as the difference between sanitized and non-sanitized CR dataset whether its in case of sensing, analysis, sharing, or mobility.} For instance, if one requests a sensing query from a dataset, then the level of privacy that an observer will see is the actual anonymity of that privacy preservation mechanism. Considering this discussion, it can be concluded that one needs to take care of anonymity level as well during development of differentially private CR mechanisms.  

\subsubsection{Computational Complexity}
\maaz{Computational complexity is a very important parameter in designing a privacy} preservation strategy~\cite{crnmatric10}. Among all privacy preservation mechanisms (such as anonymization, encryption, information-theoretic privacy.) differential privacy has minimum computational complexity due to its light-weight nature~\cite{crnfund13}. However, even during development of differentially private CR mechanisms, the aspect of computational complexity cannot be ignored. This is because of the fact that sometimes, CR nodes are pretty minute, and they cannot handle large computations. For instance,  let us take the case of a low-powered agricultural sensor which is measuring crop parametric values and is carrying out communication through CR. In this case, multiple sensors taking these readings will be pretty minute and will not have enough computational complexity to carry out heavy tasks. And if one thinks of integrating local differential privacy with these nodes, \maaz{then it will need such a local differential privacy} mechanism which is computationally efficient and can be supported by these nodes. Therefore, it is important to ensure that the complexity of differentially private CR mechanisms is well suited for the concerned application.


\begin{table*}[t!]

 \centering
 \captionsetup{labelsep=space}
 \small
 \captionsetup{justification=centering}
 \caption{\textsc{\\\comst{\cogcom{Benchmarks achieved from the perspective of differential privacy, cognitive radio networks, and their integration}}}}%
\label{taxonomycps}
{\color{black}\begin{tabular}{|r |@{\foo} l|}

\hline
\midrule
\rule{0pt}{2.5ex}
\tikzmark{a}1999 & ~\cite{crnintro05} ~ \cogcom{J. Mitola coined the concept of cognitive radio as a dynamic and intelligent radio.}\\
\rule{0pt}{2.5ex}
2006 & \cite{crnintro07} ~ \cogcom{C. Dwork introduced the notion of differential privacy to protect private data during queries.}\\

\rule{0pt}{2.5ex}
2007 & \cite{cogcomref08} ~ \cogcom{H. Celebi highlighted the need of privacy preservation in location-aware CRN.}\\
\rule{0pt}{2.5ex}
2008 & \cite{cogcomref09} ~ \cogcom{N.R. Prasad discussed security \& privacy concerns in CRN alongside proposing a secure authentication framework.} \\
\rule{0pt}{2.5ex}
2012 & \cite{crnsource08} ~ \cogcom{S. Li worked over preserving location privacy during collaborative spectrum sensing of CRN via differential privacy.}\\
\rule{0pt}{2.5ex}
2014 & \cite{survey02} ~ \cogcom{W. Wei wrote a comprehensive book on preserving location privacy for CRN.} \\
\rule{0pt}{2.5ex}
2014 & \cite{bidpriv04} ~ \cogcom{R. Zhu integrated differential privacy in spectrum trading auction of CRN. }\\
\rule{0pt}{2.5ex}
2014 & \cite{specpriv02} ~ \cogcom{W. Wang preserved CR sensing privacy leakage attacks via differential privacy in multi-service provider scenarios.}\\
\rule{0pt}{2.5ex}
2017 & \cite{survey04} ~ \cogcom{Mohamed wrote a detailed survey article on the issue of location privacy leakage in cognition cycle of CRN.}\\
\rule{0pt}{2.5ex}
2018 & \cite{locpriv04} ~ \cogcom{X. Dong introduced a differentially private notion to protect operation time privacy of primary users.}\\
\rule{0pt}{2.5ex}
\tikzmark{b} 2019 & \cite{bidpriv05} ~ \cogcom{F. Hu worked over development of differentially private matching based double auction in spectrum trading.} \\
\bottomrule
\hline
\end{tabular}}
\tikz[remember picture,overlay] \draw[->] (a.center -| b.center) -- (b.center);
\end{table*}


\subsubsection{Utility Evaluation}
\maaz{Utility is one of the most considered factors during the development of privacy} preservation models because it determines the usefulness of the mechanism. Similar is the case with differential privacy models, where utility plays a very important role in determining privacy budget and anonymity level~\cite{crnmatric11}. Similarly, in differentially private CRN models the utility also plays a very important role because this factor is responsible for smooth functioning of the network. \maaz{For instance, if the utility of SS values is low, then CR nodes shall not be able to access spectrum in the best way, which in turn shall impact the spectrum utilization. Similarly, if utility is not considered during spectrum trading then the social welfare being of auction may go negative.} Similarly, the possibility of auction being in non-equilibrium state is also there if auction utility is not considered properly. Therefore, during development of differentially private CRN models, utility needs to be considered in detail. In order to do so, researchers specifically evaluate utility parameters in their experimental evaluation to ensure that the utility of the proposed mechanism is up to a specific level, and noise addition has not disturbed utility much.

\subsection{Attack Resilience}
\cogcom{Preventing adversarial attack is one of the most prominent features of a privacy model. Different privacy mechanisms provide resilience to different types of attacks. Similarly, the differential privacy model also provides strong resilience to a lot of privacy attacks~\cite{crnmatric12}.} \maaz{Nevertheless, differential privacy provides resilience to a number of attacks. There are some attacks which need special consideration while developing differentially private CR models due to the nature of the adversary involved. In this section, three major attacks that are of sheer importance in preserving CRN privacy are discussed.}
\subsubsection{Inference Attack}
The first attack that requires special consideration is inference attack in CRN. As the name suggests, the adversary tries to infer private data of participants by carrying out various statistical analysis~\cite{crnmatric01}. Similarly, in certain cases, the adversary tries to use various machine/deep learning tools as well to find out more about the participants involved in the dataset. In CRN, this inference attack is usually carried out by an external adversary who tries to find out more information from a centralized server by asking unethical queries or by gaining maximum possible access to the database. For instance, the adversary can try to ask multiple queries related to a single person in a database to find more about that particular individual. Therefore, while developing differentially private CRN models, it is important to evaluate and check the effect of inference attacks on databases. This is usually done by asking multiple queries and evaluating the privacy leakage effect through differentially private answers. In this way, one can analyse if the proposed differentially private model is resilient to inference attack or not.
\subsubsection{Disclosure Attack}
Disclosure attack is more related to leakage of private information about a particular individual, spectrum band, or an organization~\cite{crnmatric02}. In certain cases, the organizations/FCs have to share the dataset to observers in order to perform certain statistical tasks. But they do not share the dataset directly, they first anonymize the database through a privacy preservation mechanism and then share the dataset for statistical tasks. Indeed, FCs try their best to make the private information protected, but if the data analysing observer is an adversary then it tries to carry out a disclosure attack in order to infer private information. To overcome this, researchers are not integrating the phenomenon of differential privacy before publicizing databases. Therefore, disclosure attack analysis is pretty important in cases where the possibility of publishing anonymized data is high. Differential privacy mechanism provides strong resilience to this attack, as it randomizes databases in a manner that the presence/absence of a particular individual cannot be guessed with confidence. Because the value of noise always keeps the observer in ambiguity, it cannot predict anything with confidence.
\subsubsection{Correlation Attack}
\maaz{Another critical attack that needs to be considered while developing differentially private CR models is correlation attack. A correlation attack is usually used after query evaluation. For example,} after query evaluation, the data of queries is stored at the adversary side, and then the adversary tries to carry out machine/deep learning based analysis over the collected information. The adversary combines the collected information with other publicly available datasets to find out links and correlation between participants~\cite{crnmatric03}. This correlation analysis is further used to infer private information of participating CR users, whether they are PUs or SUs. Therefore, while developing differentially private CR protocols it is important for researchers to analyse the effect of correlation attack in order to show resilience to it.

\subsection{From the Perspective of Cognitive Radio}

Alongside privacy and other matrices, certain parameters from CR perspective do also need careful analysis and evaluation while designing differentially private CR matrices. Usually, a networking protocol is evaluated on the basis of enhancement in throughput, communication overhead, and communication delay~\cite{crnmatric18}. However, in the case of CRN, it is also important to figure out that the proposed model also analyses the effect of PU activity in the proposed model. In this section, we provide a thorough discussion of why the evaluation of these parameters is important while designing a differentially private CR model.
\subsubsection{Incorporating PU Activity}
The most important parameter from the CR perspective of incorporation of PU activity in the proposed model. CR is known for its dynamic capability of allowing CR nodes to carry out communication in the presence of PUs. Therefore, evaluating the proposed model with certain PU activity is important. PU activity can simply be defined as the usage of spectrum by primary nodes. This usage can be of different types and various PU activity patterns have been developed by researchers called long-term, high, low, and intermittent. Each of these models have their own significance and cause different types of inference to CR usage. A detailed discussion about these models can be found in~\cite{crnsource20}. Since, these models pose different types of inferences to utility, usability, and privacy of the network, therefore, while developing differentially private CR models, it is important to incorporate these activities to show a broader perspective.

\subsubsection{Throughput Enhancement}
Throughput is another critical parameter that is used by researchers to evaluate effectiveness of CR protocols~\cite{crnmatric14}. Generally, throughput in wireless networks is considered as the total number of messages transmitted in a particular time interval~\cite{crnmatric19}. However, in CRN, throughput analysis is pretty vast and a number of scenarios come under throughput enhancement and analysis. For instance, the sensing of spectrum and PU nodes around CR nodes also comes under throughput, as the faster the sensing works, the faster will the throughput of the network~\cite{crnmatric14}. Similarly, the spectrum hand-off speed at the time of mobility of CR nodes is also considered as throughput. However, the important thing is that, if one wants to integrate differential privacy in these aspects then it is important to ensure that the throughput of CRN should enhance or at least do not decrease by addition of external noise. Therefore, while developing differential privacy CR models, authors are suggested to evaluate throughput aspect to ensure the speed of the network.

\subsubsection{Delay Enhancement}
Delay is usually taken in terms of the time taken to carry out some operation, however, in wireless networks it is the time that a packet takes from source to destination~\cite{crnmatric15}. This parameter is being considered by approximately every second work in the field of CRN because they want to ensure that their model is efficient enough to carry out seamless communication. But when we talk about integration of differential privacy in CRN, it becomes a more critical parameter. Because the communication is now obfuscated, and in order to get beneficial output, one has to take special care of delay caused due to perturbation. Therefore, while evaluating differentially private CR models, researchers are usually advised to evaluate this delay parameter alongside. 

\subsubsection{Communication Overhead Analysis}
Overhead caused by communication cannot be neglected while developing differentially private CR mechanisms because it plays an important role to carry out smooth communication between nodes~\cite{crnmatric13}. Typically, differential privacy works over noise, but this noise addition should not contribute overhead. If the overhead increases because of noise addition, then the overall network performance will be reduced. Therefore, during development of differentially private CR models, researchers perform overhead analysis and compare the overhead of the proposed model with previous models without differential privacy. In this way, the proposed model is analysed and approved efficiently enough for practical implementation in CRN.

\subsection{Summary}
\maaz{In this section, a comprehensive analysis of all matrices that can play a critical role during designing and development of differentially private CRN protocols have been presented.} Overall, we divide parameters in three categories from the perspective of privacy, attacks, and CR. We then sub-divide each category further to provide a much clearer picture for our readers. Firstly, from the perspective of privacy preservation, \cogcom{we first discuss how noise level and anonymity level can play the role in determining the degree of privacy. Afterwards, we highlight that how computational complexity should be evaluated at the time of designing differentially private models,} and finally we provide insights about utility evaluation of differentially private CR models.\\
Secondly, from the perspective of attack resilience, we gave a thorough investigation about top three attacks which are important to be prevented in differentially private CR models. First attack is inference attack which is caused due to harmful inference of adversary on centralized data centres or communication links. Second attack we discuss is data disclosure attack, which is more linked to data sharing, while the third attack we analysed is correlation attack in which strong machine/deep learning models are used to find out correlation between other databases and CR databases. \cogcom{Finally, in the third parametric category, we analyse parameters from CR perspective, and provide discussion about four CR parameters which should be discussed in technical works.} Firstly, we discuss primary user incorporation, then we provide discussion about throughput enhancement, afterwards, we analyse delay and overhead of CRN. Overall, it can be concluded that if one is designing a differential privacy based CR protocol, then analysing and evaluating these parameters will help shape the work in the best manner.


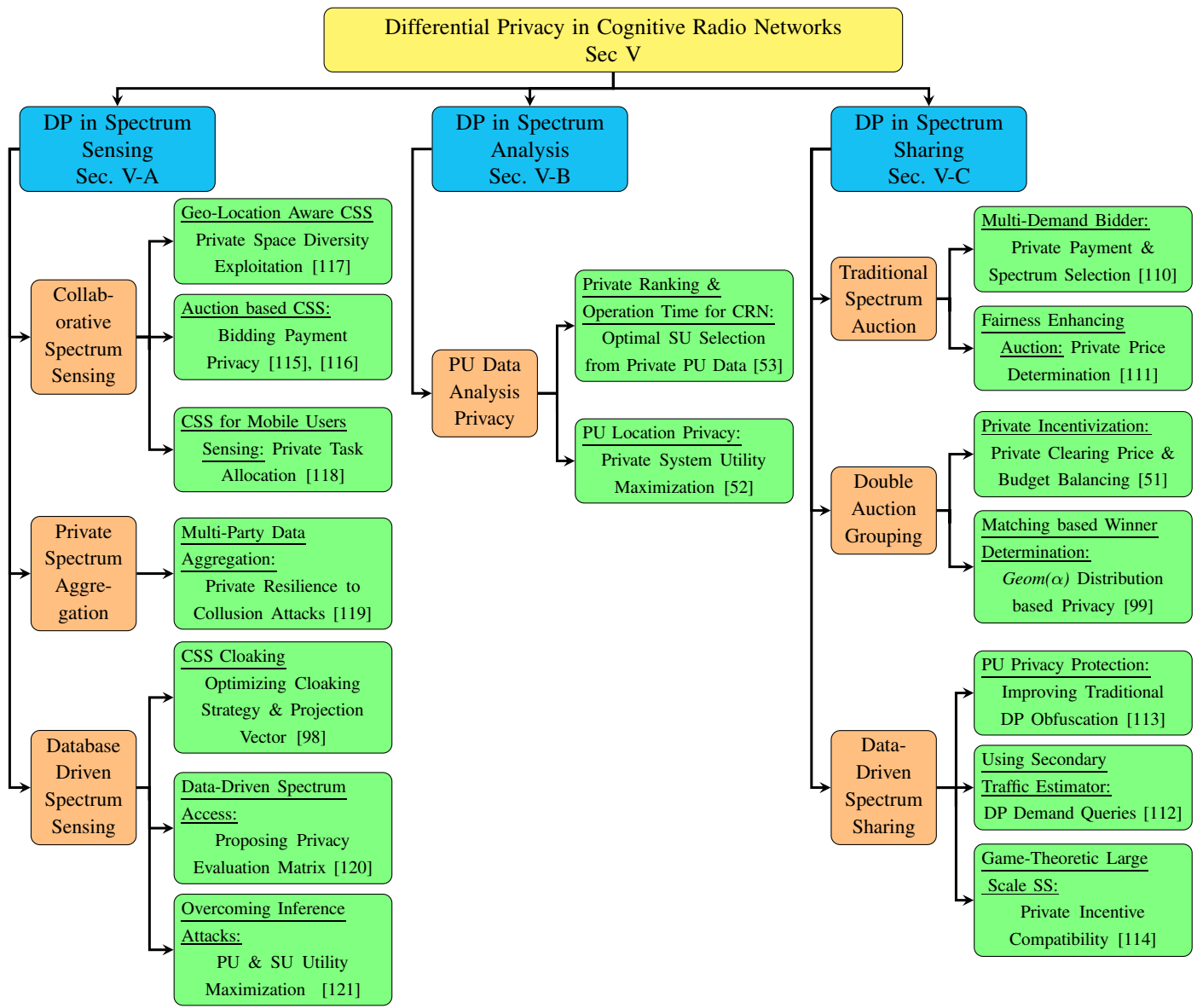
\begin{figure*}[]
\centering

\begin{tikzpicture}

\node [block,  text centered, fill=yellow!70, minimum width = 25em,  text width=25em] (a1) {Differential Privacy in Cognitive Radio Networks\\ Sec~\ref{Sec:Technical} };


\node[block, below of=a1, yshift=-2em, xshift = -3.7em, fill=cyan!70, text width=08em ](b1){{DP in Spectrum Analysis\\ Sec.~\ref{DPLocation}}};


\node[leftblock, below of=b1, yshift=-8em, xshift = -2em, text width=04em ](b1c3){{\small{PU Data Analysis Privacy}}};
\node [smallblock, right of=b1c3, xshift = 6em, yshift=3em,text width=9em] (b1c3d1) {{\footnotesize{\underline{Private Ranking \&}  \newline \underline{Operation Time for CRN:} \newline Optimal SU Selection from Private PU Data~\cite{locpriv04} }}};

\node [smallblock, right of=b1c3, xshift = 6em, yshift=-3em,text width=9em] (b1c3d2) {{\footnotesize{\underline{PU Location Privacy:} \newline Private System Utility Maximization~\cite{locpriv05} }}};





\node[block, below of=a1, yshift=-2em, xshift = 14em, fill=cyan!70,  text width=08em ](b2){{DP in Spectrum Sharing\\ Sec.~\ref{DPBid}}};

\node[leftblock, below of=b2, yshift=-3.8em, xshift = -2em, text width=04em ](b2c2){{\small{Traditional Spectrum Auction}}};

\node [smallblock, right of=b2c2, xshift = 6em, yshift= 2.2em,text width=9em] (b2c2d3) {{\footnotesize{\underline{Multi-Demand Bidder:} \newline Private Payment \& Spectrum Selection~\cite{bidpriv01}}}};
\node [smallblock, right of=b2c2, xshift = 6em, yshift= -2.2em,text width=9em] (b2c2d2) {{\footnotesize{\underline{Fairness Enhancing} \newline \underline{Auction:} Private Price Determination~\cite{bidpriv02}}}};




\node[leftblock, below of=b2, yshift=-13.2em, xshift = -2em, text width=04em ](b2c4){{\small{Double Auction Grouping}}};

\node [smallblock, right of=b2c4, xshift = 6em, yshift= 2.5em,text width=9em] (b2c4d1) {{\footnotesize{\underline{Private Incentivization:} \newline Private Clearing Price \& Budget Balancing~\cite{bidpriv04}}}};
\node [smallblock, right of=b2c4, xshift = 6em, yshift= -2.5em,text width=9em] (b2c4d2) {{\footnotesize{\underline{Matching based Winner} \newline \underline{Determination:} \newline \textit{Geom($\alpha$)} Distribution based Privacy~\cite{bidpriv05}}}};

\node[leftblock, below of=b2, yshift=-25.5em, xshift = -2em, text width=04em ](b2c1){{\small{Data-Driven Spectrum Sharing}}};
\node [smallblock, right of=b2c1, xshift = 6em, yshift=0em,text width=9em] (b2c1d1) {{\footnotesize{\underline{Using Secondary} \newline \underline{Traffic Estimator:} \newline DP Demand Queries~\cite{bidpriv06}}}};

\node [smallblock, right of=b2c1, xshift = 6em, yshift=4.2em,text width=9em] (b3c3d2) {{\footnotesize{\underline{PU Privacy Protection:} \newline Improving Traditional DP Obfuscation~\cite{specpriv08}}}};

\node [smallblock, right of=b2c1, xshift = 6em, yshift=-5.0em,text width=9em] (b3c2d1) {{\footnotesize{\underline{Game-Theoretic Large} \newline \underline{ Scale SS:} \newline Private Incentive Compatibility~\cite{specpriv03}}}};




\node[block, below of=a1, yshift=-2em, xshift = -22em, fill=cyan!70,  text width=08em ](b3){{DP in Spectrum Sensing\\ Sec.~\ref{DPSpecSen}}};

\node[leftblock, below of=b3, yshift=-5.5em, xshift = -1.5em, text width=04em ](b3c1){{\small{Collab-\\ orative Spectrum Sensing}}};
\node [smallblock, right of=b3c1, xshift = 6em, yshift=0em,text width=9em] (b3c1d1) {{\footnotesize{\underline{Auction based CSS:} \newline Bidding Payment Privacy~\cite{specpriv04,specpriv06}}}};

\node [smallblock, right of=b3c1, xshift = 6em, yshift=4.25em,text width=9em] (b1c2d1) {{\footnotesize{\underline{Geo-Location Aware CSS} \newline  Private Space Diversity Exploitation~\cite{locpriv01} }}};

\node [smallblock, right of=b3c1, xshift = 6em, yshift=-5em,text width=9em] (b1c2d2) {{\footnotesize{\underline{CSS for Mobile Users} \newline \underline{Sensing:} Private Task Allocation~\cite{locpriv03}}}};

\node[leftblock, below of=b3, yshift=-16em, xshift = -1.5em, text width=04em ](b3c2){{\small{Private Spectrum Aggregation}}};

\node [smallblock, right of=b3c2, xshift = 6em, yshift=0em,text width=9em] (b3c2d2) {{\footnotesize{\underline{Multi-Party Data} \newline \underline{Aggregation:} \newline Private Resilience to Collusion Attacks~\cite{specpriv05}}}};

\node[leftblock, below of=b3, yshift=-25.5em, xshift = -1.5em, text width=04em ](b3c3){{\small{Database Driven Spectrum Sensing}}};
\node [smallblock, right of=b3c3, xshift = 6em, yshift=4em,text width=9em] (b3c3d1) {{\footnotesize{\underline{CSS Cloaking} \newline Optimizing Cloaking Strategy \& Projection Vector~\cite{specpriv02} }}};
\node [smallblock, right of=b3c3, xshift = 6em, yshift=-1.8em,text width=9em] (b3c4d1) {{\footnotesize{\underline{Data-Driven Spectrum} \newline \underline{Access:} \newline Proposing Privacy Evaluation Matrix~\cite{specpriv10}}}};
\node [smallblock, right of=b3c3, xshift = 6em, yshift=-7.2em,text width=9em] (b1c4d1) {{\footnotesize{\underline{Overcoming Inference} \newline \underline{Attacks:} \newline PU \& SU Utility Maximization ~\cite{locpriv07}}}};


\path [line] (a1)-- ($(a1.south)+(0,-0.25)$) -|(b1);
\path [line] (a1)-- ($(a1.south)+(0,-0.25)$) -|(b2);
\path [line] (a1)-- ($(a1.south)+(0,-0.25)$) -|(b3);



\path [line] (b1c3.east) -| ($(b1c3.east)+(0.3,-0.3)$) |-(b1c3d1.west);
\path [line] (b1c3.east) -| ($(b1c3.east)+(0.3,-0.3)$) |-(b1c3d2.west);


\path [line] (b2c1.east) --(b2c1d1.west);
\path [line] (b2c1.east) -| ($(b2c1.east)+(0.3,-0.3)$) |- (b3c3d2.west);
\path [line] (b2c1.east) -| ($(b2c1.east)+(0.3,-0.3)$) |-(b3c2d1.west);

\path [line] (b2c2.east)-| ($(b2c2.east)+(0.2,-0.2)$) |-(b2c2d2.west);
\path [line] (b2c2.east)-| ($(b2c2.east)+(0.2,-0.2)$)|-(b2c2d3.west);


\path [line] (b2c4.east)-| ($(b2c4.east)+(0.2,-0.2)$) |-(b2c4d1.west);
\path [line] (b2c4.east)-| ($(b2c4.east)+(0.2,-0.2)$) |-(b2c4d2.west);

\path [line] (b3c1.east) -| ($(b3c1.east)+(0.2,-0.2)$) |- (b3c1d1.west);
\path [line] (b3c1.east) -| ($(b3c1.east)+(0.2,-0.2)$) |- (b1c2d1.west);
\path [line] (b3c1.east) -| ($(b3c1.east)+(0.2,-0.2)$) |- (b1c2d2.west);

\path [line] (b3c2.east) -- (b3c2d2.west);

\path [line] (b3c3.east) -| ($(b3c3.east)+(0.2,-0.2)$) |- (b3c3d1.west);
\path [line] (b3c3.east) -| ($(b3c3.east)+(0.2,-0.2)$) |- (b3c4d1.west);
\path [line] (b3c3.east) -| ($(b3c3.east)+(0.2,-0.2)$) |- (b1c4d1.west);





\path [line] (b1.west)-| ($(b1.west)+(-0.3,0)$) |-(b1c3.west);

\path [line] (b2.west)-| ($(b2.west)+(-0.3,0)$) |-(b2c1.west);
\path [line] (b2.west)-| ($(b2.west)+(-0.3,0)$) |-(b2c2.west);
\path [line] (b2.west)-| ($(b2.west)+(-0.3,0)$) |-(b2c4.west);

\path [line] (b3.west)-| ($(b3.west)+(-0.15,0)$) |-(b3c1.west);
\path [line] (b3.west)-| ($(b3.west)+(-0.15,0)$) |-(b3c2.west);
\path [line] (b3.west)-| ($(b3.west)+(-0.15,0)$) |-(b3c3.west);


\end{tikzpicture}

	\small \caption{A Detailed Classification of Technical Works Integrating the Concept of Differential Privacy (DP) in Cognitive Radio Networks at Various Cognitive Cycle Scenarios}
     \label{fig:tnfigTechnical}
\end{figure*}



\begin{table*}[htbp]
\begin{center}
 \centering
 \scriptsize
 \captionsetup{labelsep=space}
 \captionsetup{justification=centering}
 \caption{\textsc{\\A Parameter based Evaluation of Technical Works Integrating Differential Privacy in Cognitive Radio Networks}}
  \label{tab:techtab01}
  \begin{tabular}{|P{1cm}|P{1.4cm}|P{0.5cm}|P{3cm}|P{1.2cm}|P{1.2cm}|P{1.2cm}|P{1.2cm}|P{0.4cm}|P{0.5cm}|P{0.4cm}|P{0.4cm}|P{0.4cm}|}
  \hline

&  & & & & & & & \multicolumn{4}{c}{\centering  \bfseries ~~~~~~CRN Parameters} &\\
\cline{9-13}
\rule{0pt}{2ex}
\centering  \bfseries Domain  & \centering  \bfseries Sub-Domain & \centering \bfseries Ref. & \centering  \bfseries Major Contribution & \centering  \bfseries System Model & \centering \bfseries DP Mechanism & \centering  \bfseries Privacy Criterion & \centering  \bfseries Complexity & \bfseries PU \newline acti \newline vity & \centering \bfseries Thr \newline ough \newline put & \centering \bfseries Util \newline ity & \centering \bfseries Del\newline ay & \bfseries Over \newline head\\

\hline

\multirow{5}{*}{\parbox{2cm}{\centering \textbf{}}}
 &  & ~\cite{specpriv04} & Private location \& payment strategy during crowdsensed spectrum sharing. & Centralized & Exponential & $\varepsilon$-dp & $-$ & \centering \tick & \cross & \centering \tick & \cross & \tick \\
\cline{3-13}

&  & ~\cite{specpriv06} & \newmod{Enhanced bidding privacy} for multi-bid CSS users. & Centralized & Exponential & ($(e-1)\varepsilon \prime \newline \delta ln(e \delta^{-1}), \newline \delta$)-dp & $-$ & \cross & \cross & \centering \tick & \cross & \tick \\
\cline{3-13}

 & \newmod{Collaborative Spectrum Sensing} & ~\cite{locpriv01} & Preserving Geo-location for CRN participants in collaborative network. & Centralized \newline (cluster based) & Manual & ($\mu,\delta$)-dp & $\lceil{\frac{n}{m}}\rceil . $ \newline $(\frac{m+1}{2}).\newline C_{tr} $  & \cross & \cross & \cross  & \cross  & \tick  \\
\cline{3-13}

&  & ~\cite{locpriv03} & \newmod{Private location-aware geocast SS for CRN.} & Centralized & Laplace & $\varepsilon$-dp & $-$ & \cross  & \cross & \centering \tick & \cross & \tick  \\
\cline{3-13}

\cline{2-13}

\centering \textbf{Spectrum Sensing} & Private Spectrum Aggregation & ~\cite{specpriv05} & Spectrum aggregation \& auction for multi-party CRN. & Centralized & Manual & $\varepsilon$-dp & $O(nn_i) \newline Add$ & \cross & \cross & \centering \tick & \cross & \tick \\
\cline{2-13}

&  & ~\cite{specpriv02} & Private collaborative sensing for service providers privacy via Cloaking Time. & Distributed \newline Clusters& Laplace & $\varepsilon$-dp & $O(N^2K)$ & \centering \tick & \cross & \centering \tick & \cross & \tick \\
\cline{3-13}

& \newmod{Database Driven Private Sensing} & ~\cite{locpriv07} & PU \& SU utility maximization via Optimal Private Decisions. & Centralized & Laplace & $\varepsilon$-dp & $-$ & \centering \tick & \cross & \centering \tick & \cross & \cross \\
\cline{3-13}

&  & ~\cite{specpriv10} & Evaluating Collaborative Privacy Protection Models of Spectrum Access Systems and PU. & Centralized & Exponential & $\varepsilon$-dp & $-$ & \cross & \cross & \centering \tick & \centering \tick  & \tick \\

\hline

\multirow{2}{*}{\parbox{2cm}{\centering \textbf{}}}

\centering \textbf{Spectrum Analysis} & PU Data Analysis Privacy & ~\cite{locpriv04} & Selecting Optimal SUs from Private operation-time values for CRN. & Centralized & Exponential & ($2 \varepsilon \Delta$)-dp & $O(ln(N))$ & \centering \tick  & \cross & \centering \tick & \cross & \cross  \\
\cline{3-13}

 &  & ~\cite{locpriv05} & Protecting PU location privacy optimally. & Centralized & Exponential & ($w,\varepsilon$)-dp & $-$ & \cross  & \centering \tick & \centering \tick & \centering \tick & \cross  \\
\cline{2-13}

\hline

\hline
\multirow{8}{*}{\parbox{2cm}{\centering \textbf{}}}
 & Traditional Spectrum Auction & ~\cite{bidpriv01} & Private payment \& spectrum selection for multi-demand bidder. & Centralized & Exponential & ($2 \varepsilon \Delta$)-dp & $-$ & \cross & \cross & \centering \tick & \cross &  \cross \\
\cline{3-13}
& & ~\cite{bidpriv02} & Fairness enhancing private auction. & Centralized & Exponential & ($2 \varepsilon \Delta q$)-dp & $-$ & \cross & \cross & \centering \tick & \cross & \cross  \\
\cline{2-13}


\centering \textbf{Spectrum Sharing} & Double Auction Grouping & ~\cite{bidpriv04} & Private incentivization and budget balancing for CRN. & Centralized & Exponential & $\varepsilon$-dp & $-$ & \cross & \cross & \centering \tick & \centering \tick & \cross \\
\cline{3-13}
 &  & ~\cite{bidpriv05} & \textit{Geon($\alpha$)} distribution based bid matching privacy. & Centralized & Laplace & $\varepsilon$-dp & $O(KM^2 \newline N^2X)$ & \cross & \cross & \centering \tick & \cross & \cross \\
\cline{2-13}

&  & ~\cite{bidpriv06} & Private demand query modelling using secondary traffic estimator. & Centralized & Laplace & $\varepsilon$-dp & $-$ & \cross & \cross & \centering \tick & \cross & \cross \\
\cline{3-13}

& \newmod{Data-driven Spectrum Sharing} & ~\cite{specpriv03} & Truthful game-theoretic aggregation for large scale spectrum sharing. & Centralized \newline (cluster based) & Exponential & $\varepsilon$-dp & $-$ & \cross & \cross & \centering \tick & \centering \tick & \cross \\
\cline{3-13}
&  & ~\cite{specpriv08} & Protecting PU privacy in spectrum sharing. & Centralized & Exponential & ($\varepsilon, \delta$)-dp & $-$ & \centering \tick & \centering \tick & \centering \tick & \cross & \tick \\
\cline{2-13}

\hline

 \end{tabular}
  \end{center}
\end{table*}


\section{Differential Privacy Approaches for Cognitive Radio Networks} \label{Sec:Technical}

\maaz{In previous sections, we discussed how differential privacy can play its role in developing private CRN models, and that parameters should be considered while developing differentially private CR protocols. In this section, an extensive literature review that involves integration of differential privacy with CRN at different models is discussed. Based on the cognitive cycle, the technical work has been divided into three three categories, that is differential privacy in: (i) SS, (ii) spectrum analysis, and (iii) spectrum sharing. Evaluation of the technical work is given in Table~\ref{tab:techtab01}, whereas classification of the technical work is given in Fig.~\ref{fig:tnfigTechnical}.} 

\subsection{Differential Privacy in Spectrum Sensing}\label{DPSpecSen}
Spectrum sensing involves carrying out measurement and observations for efficient utilization of spectrum bands~\cite{crntech01}. In SS, SUs sense their surroundings and develop a radio environment map by sharing these values with each other to get better results~\cite{crntech02}. \maaz{The sensing could be through centralized FC (which is also known as database-driven SS) or it could be through collaborative or crowdsourced approach in which all CR nodes collaborate to figure out the optimal environment. No matter which approach is used; the risk of privacy leakage still exists.} For instance, location privacy, identity privacy, and behavioural privacy of PUs, SUs, and FC is always at risk when multiple parties are involved. Therefore, it is important to integrate a privacy preservation mechanism in order to protect privacy in an efficient manner. \maaz{We now discuss technical approaches which propose this integration and develop their models.}
\subsubsection{\mubashir{Collaborative Spectrum Sensing}}
\mubashir{Sensing spectrum in a collaborative fashion is one of the prominent steps of CRN, as this gives information about surroundings in order to access the most beneficial spectrum. However, this step is vulnerable to privacy leakage as well, and to overcome this privacy leakage, differential privacy can play a viable role. The first work in this direction providing} detailed insights about the development of a differentially private incentive mechanism has been presented by Dong~\textit{et al.}~\cite{specpriv06}. The \cogcom{work developed a CSS model and integrated the concept} of differential privacy to protect privacy. Authors developed a multi-bid model which collects bids from sensing participants and selects sensing winners in differentially private manner. In the proposed work, authors ensured that they maximize social welfare alongside enhancing winners’ cost efficiency. \maaz{The work provided extensive theoretical analysis for differential privacy guarantees, however, an analysis from the perspective of welfare maximization is missing the article.} Another work discussing differential privacy for crowdsourced SS of CRN have been carried out by Jin~\textit{et al.}~\cite{specpriv04}. Authors proposed two models named as PriCSS$^+$ and PriCSS$^+$ and demonstrated that both of the proposed models preserve location privacy using differential privacy protection. Alongside protection, authors also enhanced payment system and proposed a truthful and cost minimizing payment mechanisms of SS participants. From the perspective of evaluation, authors carried out evaluation on multiple PU activities and evaluated utility and network overhead to demonstrate that their proposed model outperforms others in this perspective.\\
\maaz{A similar work on location privacy preservation for CSS is presented by Li~\textit{et al.}~\cite{locpriv01}. Authors proposed a privacy preserving sensing scheme,PPSS. The authors further} proposed two primate protocols of PPSS for aggregation and injection during sensing called PPRSA and DDRI respectively. Authors evaluated the proposed model on SRLP and DLP attacks by taking values from real-testbed sampling region of CRN. From experimental evaluations, it can be seen that the proposed model successfully protected privacy of model alongside enhancing entropy and fluctuations of received signal strength. The fourth work in the domain of differentially private privacy protection for crowdsourced SS have been carried out by Huang and Gong~\cite{locpriv03}. The work featured the method of geo-cast for hop by hop message dissemination and broadcast in CRN, and in the article authors demonstrated that how location privacy gets leaked via this type of sensing. The proposed model enhanced privacy protection for geo-cast SS alongside providing enhancement in agents task acceptance, system overhead, and report correlation. \maaz{In the experimental evaluation, the work analyzed correlation in the data, however, the aspect of privacy leakage and error rate is missing, which is one of the critical aspect that should be added in the location reporting works.}
\subsubsection{\mubashir{Private Spectrum Aggregation}}
\mubashir{Apart from CSS, a work discussing private sensing aggregation and auction in CRN is presented by Zhou~\textit{et al.}~\cite{specpriv05}.} The work first proposed an efficient data aggregation model for multiparty CRN, and then demonstrated the sources of privacy leakage in this model. Thus, in order to overcome the privacy leakage, the authors further propose a light-weight privacy preserving aggregation strategy using the concept of dynamic differential privacy. And, alongside providing private aggregation, authors also proposed PPSSA for differentially private spectrum auction to enhance revenue of auctioneer. The authors further evaluated their proposed models and experimental evaluation showing that the proposed strategy not only enhances communication and communication cost, but also enhanced SUs satisfaction and auctioneers’ revenue. \maaz{Authors provide an in-depth theoretical analysis for the proposed work; however, it is important to mention that the article does not have any algorithm, which is usually helpful in replicating the work for future experiments by new researchers.}
\subsubsection{\mubashir{Database Driven Spectrum Sensing}}
\mubashir{In SS for CRN, one cannot ignore the discussion about database driven SS, which is the second most prominent strategy to carry out SS after CSS~\cite{crnsource36}. This method is dominantly used by SUs to get efficient results, but research has indicated that it can also cause privacy leakage. In order to overcome this privacy leakage, researchers have proposed usage of differential privacy in database driven SS.} \maaz{The first work in the integration of differential privacy in CSS have been carried out by Wang~\textit{et al.}~\cite{specpriv02}. It is a privacy preserving framework for SU in CRN on basis} of cloaking time and named the framework as PromCos. Authors draw the motivation that in order to carry out efficient sensing, it is important to incentivize and enhance trust of SUs in the network. In order to do so, authors computed differentially private theoretical privacy guarantees for SU, and afterwards, evaluated these guarantees on SP, SU, and collusion attack to ensure privacy.
\maaz{Some other works evaluated database driven private sensing from a differentially private perspective. The first work in enhancing privacy of database driven CRN was carried out by Zhang~\textit{et al.}~\cite{locpriv07}. The aim of the article is to develop an inference free framework for both SUs and PUs in which both will be able to play their part in SS without the risk of losing their private location and identity data.} The author formulated theoretical bounds for optimal decisions of PUs and SUs, and then evaluated these bounds by carrying out extensive experimental evaluation, which ensured utility of both participating SUs and PUs. The final work for integration of differential privacy in CRN was carried out by Li~\textit{et al.}~\cite{specpriv10}. The work focused over evaluation of privacy preserving models for SU privacy in database driven CRN. The proposed work analysed four different types of adversaries and then proposed two security matrices called indistinguishable input and adversarial estimation error. Afterwards, the author evaluated the privacy preservation models and showed the effect of these models with respect to proposed matrices. \maaz{Overall, it will not be wrong to say that the article is a good piece of literature for a new researcher in the field. But in-depth analysis, evaluation, and experiments from the perspective of the field are lacking in the article.}

\subsection{Differential Privacy in Spectrum Analysis}\label{DPLocation}
Analysing spectrum efficiently after getting values from SS devices is the second and one of the most critical steps in cognitive cycle, because this step is used to opt or bid for a specific spectrum or not~\cite{crntech03}. It also checks if a specific available spectrum is suitable for the required communication. Therefore, a careful analysis of spectrum is carried out in this step which involves two major steps named character analysis and parameter configuration. However, careful analysis has shown that the privacy of participants, especially the privacy of PUs gets leaked during this step~\cite{crntech04}. \cogcom{Therefore, it is important to protect the privacy of participating nodes alongside providing efficient service. In order to protect the privacy of participating PUs and SUs during spectrum analysis, we believe the obfuscation mechanism of differential privacy can be utilized in an efficient manner.}
\subsubsection{\mubashir{Primary User Data Analysis Privacy}}
\maaz{From the \mubashir{perspective of integration of differential privacy in spectrum analysis, it can be observed that work has only been carried out to protect PU privacy. In this direction, two significant} works have been carried out so far.} Both focus on preserving privacy of participating PUs during spectrum analysis, as PUs are the most vulnerable participants because their data is being analysed in this step. The first work for protecting operation-time privacy of PU in CRN is carried out by Dong~\textit{et al.}~\cite{locpriv04}. The authors named the proposed mechanism as PriDSS, in which they carried out a two-fold contribution. First from the perspective of providing efficient selection of SUs for dynamic sharing systems, and secondly from the perspective of providing privacy on operation time of PUs. This work integrated and contributed to multiple aspects of spectrum analysis within a single article. For instance, they worked on a SU selection problem by calling it a utility dependent problem, and then solved it by proposing a utility maximizing model. Afterwards, they explicitly mentioned that the operation-time of participating PUs is not private, and it can be used for various malicious purposes. For instance, detecting presence/absence of a particular PU or channel. After this discussion, they propose the strategy to efficiently select SUs from differentially private operation-time readings. \maaz{The evaluation results of the proposed model showed that the work efficiently overcome privacy loss alongside providing efficient payments to maximize system utility.}\\
\cogcom{The second and the final work in the domain of differentially private PU privacy protection for CRN is carried out by Liu~\textit{et al.}~\cite{locpriv05}. The article focused on protecting the privacy of PUs in a real-time environment. The major motivation of the article is the real-time protection, as the authors mentioned that several other works protected privacy in database-driven environments, but the works protecting privacy in real-time are missing.} Therefore, authors proposed a PU privacy protection model, which protects location privacy of PUs during spectrum analysis in a real-time environment. The article used the concept of cloaking time optimality to design a differentially private utility optimal model which also enhances spectrum usage efficiency. Alongside this, authors proposed theoretical guarantees \maaz{to prove that how} their proposed model obeys the rules of differential privacy in order to protect PUs from harmful adversarial interference. \cogcom{To demonstrate it further, they developed the notion of `expected interference error’ and evaluated and compared this notion with other recent works to show the significance of the proposed model.  Nevertheless, there is not much work in the domain of differentially private spectrum analysis, but the two works demonstrated how differential privacy can enhance privacy in this step optimally.} Therefore, there is a need for more work from the perspective of differentially private spectrum analysis.

\subsection{Differential Privacy in Spectrum Sharing} \label{DPBid}
Spectrum sharing is a decisive step in the cognitive cycle because multiple SUs are competing for an available spectrum at this stage. Usually, auctions are carried out to choose the optimal combination of buyer and seller which maximizes utility of auction~\cite{crntech05}. These auctions are carried out in multiple different ways. For instance, some research work is on traditional auction styles, while some proposed crowdsensed auctions. Similarly, certain other works highlighted the use of auction grouping for CRN nodes on the basis of the double auction model. Among all these auctions, it is made sure that the proposed model is rational, truthful, and maximizes social welfare of the network~\cite{crntech06}. Apart from auction models, another aspect of data analysis during spectrum sharing cannot be ignored, because it is used to learn and adapt efficient futuristic models of spectrum sharing~\cite{survey04}. Usually, the aspects of game-theory and machine/deep learning are used at this step to find optimal sharing strategies. In all these models, it is important to mention that the risk of privacy leakage cannot be ignored, and in order to overcome this privacy risk, the aspect of differential privacy is being integrated by researchers.\\ In this section, we highlight the works which integrate the concept of differential privacy in spectrum sharing to produce optimal private results. \maaz{Differentially private spectrum sharing work can be categorized into} \mubashir{three domains from the perspective of auction model and spectrum sharing. The first two categories focus over integration of differential privacy in various types of auction, while the third category focus over integration of differential privacy in other processes involving spectrum sharing.}
\subsubsection{\mubashir{Traditional Spectrum Auctions}}
The first two works in this domain worked over integrating traditional auction models with differential privacy to ensure privacy in the trading. The first work discussing revenue maximization via approximate privacy guarantees is presented by Zhu~\textit{et al.}~\cite{bidpriv01}. The work proposed a differentially private revenue maximizing auction and named the proposed strategy as DEAR. The proposed DEAR algorithm is an auction model which protects the privacy of participating entities via the Exponential mechanism of differential privacy. The proposed work introduced differentially private randomness in bidders’ group in a hexagonal formation to ensure maximum privacy. Afterwards, the work also proved fairness enhancing theoretical guarantees of differential privacy to demonstrate that their work supports all theoretical bounds. \maaz{Authors performed extensive experiments showing that the proposed work enhanced revenue of the auction market, however, it is also significant to mention that the evaluation from the perspective of social welfare of buyers is missing.} Another work that evaluated differentially private auction mechanisms in CR settings was carried out by Wu~\textit{et al.}~\cite{bidpriv02}. Authors proposed a private auction model named ‘DIARY’ and then evaluated and compared the proposed model with four other state of the art models to show effectiveness of the proposed model. It can be seen from experimental results that DIARY outperforms other models from the perspective of seller, and auction revenue. \maaz{Authors further supported their work by adding extensive theoretical contributions from the perspective of differential privacy and auction based evaluations such as winner selection and price determination.}
\subsubsection{\mubashir{Grouping based Double Auction}}
\mubashir{Another type of auction that is commonly used in differentially private CRN is grouping based double auction. \cogcom{From this perspective,} Zhu~\textit{et al.}~\cite{bidpriv04} proposed DDSM mechanism which is a double auction scheme for differentially private spectrum grouping. Authors used the Exponential mechanism of differential privacy to select private optimal prices during the double auction process.} Afterwards, authors proved using theoretical guarantees that their proposed price selection model provides truthfulness and rationality alongside providing private pricing. The second work in the domain of differentially private auction was carried out by Hu~\textit{et al.}~\cite{bidpriv05}. The authors used $Geom(\alpha)$ distribution to protect bid privacy during the auction process. The new distribution is combined with a differentially private noise addition model to find out the optimal amount of noise for the bid value before encryption. The performance evaluation demonstrated that the proposed model outperforms other similar works in terms of revenue, satisfaction, and privacy of PUs and SUs. \maaz{Since the article is a twofold contribution form perceptive of auction and differential privacy. But the experimental evaluation only shows the evaluation of auction properties, while the evaluation for privacy parameters is missing.}

\subsubsection{\mubashir{Data-Driven Spectrum Sharing}}
Apart from auction works, some researchers worked over integration of differential privacy in data-driven aspects of private spectrum sharing. The first work that falls in this category is carried out by Wang~\textit{et al.}~\cite{bidpriv06}, which focused on preserving SUs privacy in data-driven spectrum trading. The contributions of the articles are two-fold, firstly, authors preserved revenue of PUs by integrating differential privacy in it, and afterwards, authors preserved the demand values of SUs as well through dynamic differential privacy. The work first proposed an architecture in which available spectrum of PUs is aggregated for selling during spectrum sharing. PUs can sell the spectrum to centralized SSP at fixed price or can also sell it directly to SUs at the individual price. However, in both of the cases authors ensured \cogcom{that the privacy of PUs gets protected by adding a specified level of obfuscation in the trading. Alongside this, the centralized PSP also collects values and estimates the demand of SUs, in order to develop an efficient utilization matrix.} This is also done in a differentially private manner to protect privacy of SUs. Overall work ensured that the proposed model is optimized \maaz{with respect to revenue-maximization} and risk-minimization.\\
Another work targeting truthful aggregation via game-theoretic perspective is carried out by Zhou~\textit{et al.}~\cite{specpriv03}. This work introduced a novel game type in CRN and called the game as an aggregative game, which is used to carry out large-scale modelling of spectrum sharing in CRN. \cogcom{Authors draw the motivation of proposed game by saying that in large-scale CRN, the information about each other is incomplete, which forms a `weak mediator’, and this weak mediator will not be able to reach Nash equilibrium with this incomplete information. Therefore, they developed an aggregative game to improve the utility of these systems.} However, they showed that this type of learning can lead to privacy leakage, which they further eliminated by using the concept of differential privacy and proposed an incentive-compatible and differentially private approximate Nash equilibrium of aggregative game for CRN. The final work focusing over privacy enhancement of PUs in spectrum sharing has been carried out by Clark~\textit{et al.}~\cite{specpriv08}. The work first studied the trade-off between privacy and performance in differentially private CRN, and then proposed a generalized spectrum sharing model to overcome this trade-off. \cogcom{Authors considered this research gap as an optimization problem of privacy in which one measures the level of privacy on the basis of data exposure to possible adversaries. After that, the authors tried to fill this research gap by developing an optimal solution for multi-utility spectrum sharing model in terms of performance and efficiency of using spectrum in the best possible way.}

\subsection{Summary}
In this section, a comprehensive overview of integration of differential privacy in CRN from technical works perspective is presented. We first categorised all technical works from the perspective of steps involved in cognitive cycle, and afterwards provide in-depth subcategories of these cycle steps on the basis of technical contributions in the article. From the discussion it can be seen that a variety of scenarios in spectrum analysis have been evaluated ranging from \cogcom{CSS to database-drive SS. Similarly, the aspect of differentially privacy spectrum sharing has also been explored a lot from trading and database analysis perspective.} But, when we move to spectrum analysis and spectrum mobility, not much work can be seen. For instance, in spectrum analysis, only two technical works have carried out integration of differential privacy, and that too to protect PUs privacy. Similarly, from the perspective of differentially private spectrum mobility, none of the technical work evaluated this aspect in detail. Therefore, we did not provide this cycle step in the classification figure and table. This discussion opens a wide-range of future directions especially from the perspective of spectrum analysis and spectrum mobility in CRN.

\section{Applicability of Differential Privacy in Futuristic Cognitive Radios} \label{Sec:Applicability}

\rehmani{Apart from the privacy leakage in traditional usage, the notion of CR is also being applied to various applications. For instance, CRN are being integrated with smart grid and other similar technologies to carry out resource efficient communication~\cite{crnapp01}. Nevertheless, these integrations provide a vast number of benefits, but they also come with certain drawbacks, and one of the largest among them is the privacy leakage. Therefore, preserving privacy is of sheer importance in futuristic CR based technologies. In this section, discuss the importance of privacy in these applications, and then we provide an in-depth discussion that how differential privacy can be used to project privacy in these scenarios.}



\subsection{Cognitive Radio Based Smart Grid}
\rehmani{Integration of CR with traditional energy grids is also one of the key features being considered during development of modern smart grid systems~\cite{crnapp01}. This integration has provided a wide range of benefits ranging from power reduction to low-latency communication between smart meters and grid utilities~\cite{crnapp02}. For instance, CR-based smart metering nodes can carry out communication over available spectrum bands without having heavy license cost. Similarly, the highly adaptive nature of CR communication is suitable to carry out almost every possible operation of smart grid, whether its real-time reporting, fault monitoring, or meter firmware updates, etc. Nevertheless, CR-based smart grid is a viable communication model for modern smart grid scenarios, but it also comes up with certain privacy related issues which needs to be tackled. For example, daily lifestyle privacy of residents can be leaked if smart meters report their real-time data to grid utility via dynamic CR channel. Similarly, location privacy of smart meters can be leaked while carrying out spectrum analysis for communication. Therefore, it is important to protect the privacy of CR-based smart grid users before practical application of these scenarios.}
\subsubsection{Integration of Differential Privacy}
\rehmani{In order to protect privacy of CR-based smart grid users, differential privacy can be used as a viable solution because of its dynamic and light-weight nature. For instance, CR-based smart meter nodes reporting their real-time data via unlicensed channels can use a perturbation mechanism of differential privacy to add noise in their values in order to protect their privacy. Similarly, Exponential perturbation of differential privacy can be used during spectrum analysis of CR-based smart meters. Apart from these prominent directions, differential privacy can also be used to protect usage privacy during firmware updates, fault detection, fault reporting, and other similar scenarios where smart grid nodes have to use functionalities of CRN to carry out communication.  From the perspective of literature review in this domain, certain works (such as ~\cite{crnsource09, crnchallenge15}, etc.) highlighted the use of differential privacy in smart grid reporting and other analysis. However, a full-fledged work discussing and integrating differential privacy in CR-based smart grid has not been carried out yet. Therefore, we believe that this domain has potential and this integration of differential privacy with CR-based smart grid communication can lead to development of secure and efficient modern energy infrastructures.}

\subsection{Software Define Networks based Cognitive Radio}
\rehmani{Since the emergence of CR, it has been incorporated with various technologies, and software defined networks (SDNs) is one of them. This integration of CR and SDNs are in discussion for more than a decade now due to its tremendous benefits~\cite{crnapp03}. SDNs provide CR users with an efficient architecture to carry out their communication and networking needs, which in turn will help to improve the efficiency and latency of CRN. Effective network management capabilities of SDN enables CR nodes to carry out efficient communication without worrying about network latency and complexity issues. However, despite these tremendous advantages, it can also be visualized that even SDN-based CR do also suffer with privacy issues because of centrally managed networks. As SDN uses various rules to manage network wide traffic, and all this is happening in a centralized manner, therefore, if an adversary gets control over this central entity or in case of some malicious query evaluation, the privacy of these networks is at risk~\cite{crnapp04}. Therefore, it is important to protect the privacy of these networks before integrating them with CR and other similar radios.}
\subsubsection{Integration of Differential Privacy}
\rehmani{To overcome this issue of centralized query evaluation for SDN-based CRN, differential privacy serves as one of the most suitable options. Exponential mechanism of differential privacy can be used to protect privacy of CR users during query evaluation during spectrum analysis and other cognition cycles. Similarly, critical information of the SDN-based CRN can also be protected via direct data perturbation/Laplace mechanism of differential privacy in order to reduce adversarial risk. For instance, during SS when every CR node transmits their data to a centralized SDN server, then instead of reporting plain-text data, one can report differentially private data in order to protect its location and identity privacy.}

\subsection{Cognitive Radio Based Internet of Things}
\rehmani{\cogcom{IoT is a widespread domain which is playing a very important role in our daily lives ranging from small temperature sensors} in homes to massive ubiquitous sensing controllers taking decision of transportation~\cite{crnapp05}. If one looks around, IoT devices are everywhere around us and are being integrated with almost all major communication technologies to carry out operations in a streamline manner. As a part of this integration, researchers also integrated and used the concept of CR to carry out IoT communication, which has been proved to be beneficial in many ways~\cite{crnapp07}. For instance, smart in-home applications are being made capable of running over unlicensed spectrum in order to overcome spectrum scarcity due to immense increase in in-home sensors~\cite{crnapp06}. Similarly, smart cities are being developed with an aim of developing an eco-friendly environment, and one step in them is to carry out communication via CR instead of traditional licensed bands to reduce the excessive use of licensed spectrum~\cite{crnapp08}. \cogcom{Despite these tremendous benefits,} the aspect of privacy leakage from these IoT devices cannot be ignored due to the large amount of data traversing across the network~\cite{crnapp09}.}
\subsubsection{Integration of Differential Privacy}
\rehmani{IoT devices transmit a large amount of data to the network via CR unlicensed channels to carry out various operations for the functioning. If this data is not secured, then this can leak privacy of the device owners, which can lead to catastrophic results~\cite{crnapp11}. In this perspective, differential privacy can play a key role in both active and passive data protection of CR-based IoT devices. For example, in case if IoT devices are carrying real-time streaming of their data via CRN, then noise addition mechanism of differential privacy can protect data of this real-time streaming by injecting i.i.d noise in it. Similarly, in case of passive data protection for IoT where surveys, learning, and query evaluation is carried out, the randomization mechanism of differential privacy can play a role to produce ambiguity in attackers mind regarding presence or absence of a specific user or IoT node during learning. Therefore, we believe integrating differential privacy with IoT devices operating over CRN can be a viable solution to protect them from external and internal intruders.}

\subsection{\cogcom{Machine/Deep Learning based Cognitive Radio Networks}}
\cogcom{Since the advent of CRN, researchers are continuously trying to develop and integrate novel technologies to enhance its efficiency. One such pathway is integration of machine/deep learning with CRN. CRN by its basic nature is considered to be an artificially intelligent radio which is capable of taking intelligent decisions involving spectrum sharing, hand-off, etc. However, the integration of modern machine/deep learning models with CRN have opened a plethora of research directions via which one can enhance the functioning of CRN in an efficient manner~\cite{cogcomref03}. For example, reinforcement learning provides CR users the functionality to develop their behaviour via interacting with the surrounding environment during the cognition cycle~\cite{cogcomref01}. Apart from reinforcement learning, basic machine learning algorithms such as support vector machine (SVM), homomorphic machine learning, nearest neighbor, ANN, etc have also aided a lot in development of modern CRN. Similarly, in the modern times, the integration of deep learning models with CRN is paving path for future intelligent radios, e.g., one such example is the integration of deep reinforcement learning with CRN, which enhances the capability of SS a bit further and provides CRN users with an efficient signalling results to choose the best possible option~\cite{cogcomref02}. Similarly, deep neural network (DNN) based models are also being used to enhance the detection accuracy during spectrum sending~\cite{cogcomref04}.}
\cogcom{Nevertheless, the research in machine/deep learning based CRN is improving day by day, and a large amount of data from CR nodes is being collected to enhance the cognition cycle of CRN. This data is basically the feed to machine/deep learning models, and it will not be wrong to say that more data means high detection accuracy. However, on the other hand this data can leak a huge amount of private information of CR users. E.g., accurate location information during SS can lead to location privacy leakage. Similarly, spectrum hand-off information can lead to leakage of holding time privacy of nodes. Therefore, alongside collecting this huge amount of data for machine/deep learning models, it is equally important to protect this data from all sorts of adversaries.}
\subsubsection{\cogcom{Integration of Differential Privacy}}
\cogcom{Nonetheless, the collected data during the cognition cycle can pose a huge privacy concern for CR users, but on the other hand differential privacy protection can be used to protect this privacy leakage to a large extent. Nowadays, differential privacy is a universally accepted matrix for privacy protection during machine/deep learning scenarios, thus differential privacy is being integrated with almost all state-of-the-art learning models to protect privacy at different levels. For instance, local differential privacy is being used to protect individual privacy before feeding the data to a machine learning model~\cite{cogcomref05}. Similarly, differentially private learning in wireless big data networks is also being employed by researchers to test its efficacy in wireless scenarios~\cite{cogcomref06}. Apart from these, differential privacy have proved its significance in almost all machine learning models, such as naïve bayes, linear regression, linear SVM, logistic regression, kernel SVM, decision tree, k-means clustering, etc~\cite{cogcomref07}. Thus, one can imply all these differentially private machine learning models in the majority of CR scenarios. E.g., during machine learning for SS, local differential privacy can be used to protect user privacy. Similarly, during query evaluation and learning steps of cognition cycle, noise can be calibrated and added to protect individual privacy. Therefore, by keeping in view this discussion, it will not be wrong to say that differential privacy is one of the most viable solutions to protect privacy during machine learning based CR applications, where users have to share their private data in order to get efficient outcomes.}

\subsection{Cognitive Radio Based UAV Communication}
\rehmani{Nowadays, drone technology (also known as unmanned aerial vehicle (UAV)) is booming, and various different sized drones are being used to carry out multiple applications around us ranging from sports streaming to carrying critical military operations~\cite{crnapp12}. In order to carry out efficient communication and message exchange between UAVs, researchers are using unlicensed CR-bands for communication~\cite{crnapp13}. In this way, UAVs can exploit the unused spectrum and can carry out their operations without causing their part in spectrum scarcity. Nonetheless, this integration is fruitful, but research has highlighted that the information exchange via this integration is not completely secure and is prone to many privacy attacks. \cogcom{For example, in autonomous UAVs, route planning is carried out, but this route planning also needs to be reported to neighbouring clusters via CR network to overcome possibility of any collusion. However, if this information gets leaked, it can reveal the complete planning and movement of UAV clusters. Therefore, it is important to protect the privacy of these UAVs in order to carry out seamless operations.}}

\subsubsection{Integration of Differential Privacy}
\rehmani{Similar to CR-based IoT devices, CR-based UAVs are also reporting their real-time planning, location, and identity values to the neighbouring nodes, clusters, and managing authorities to carry out operations in a streamline manner. This real-time reporting can cause serious privacy leakage because significant strategic information can be leaked from the reported values. Similarly, while performing steps involved in the cognition cycle, these UAVs also have to carry out SS, analysis, sharing, and mobility, which can also cause privacy leakage to a greater extent. In order to overcome these issues, the notion of differential privacy can play a critical role due to its dynamic nature of privacy protection during a real-time reporting environment. Different from other privacy preserving strategies, differential privacy can perturb values in real-time depending upon the allowed error rate. This real-time perturbation can be integrated with UAV real-time reporting and sensing to protect their private values. For example, if a UAV is reporting its identity in real-time just to show its presence, then its identity can be protected by adding i.i.d noise, which will not affect normal operations of the CR network, but on the other hand it will protect the identity theft of that particular UAV. Therefore, we believe integrating differential privacy with CR-based UAVs can be a good step ahead to develop a secure CR-based drone network.}

\subsection{Cognitive Radio Based Industrial Internet of Things}
\rehmani{Carrying out industrial operations via industrial sensors is a well-developed sub-field of IoT which is also known as industrial IoT (IIoT). The sensors involved in IIoT are responsible to carry out decisions on the basis of received input~\cite{crnapp14}. For example, a sensor in a car manufacturing industry will take input from the buyer and will change the color of the car according to the requirement. In order to provide seamless communication, these sensors are now being operated over unlicensed CR bands~\cite{crnapp15}. This integration of CR with IIoT architecture is enabling many industries and organizations to reduce their contribution in spectrum scarcity. However, on the other hand, these IIoT nodes are now more vulnerable to privacy issues because now they have to report their sensing values to the centralized CR moderators in order to perform steps involved in cognition cycle. Therefore, it is important to protect privacy of these CR-based IIoT nodes so that they can carry out their operations without risking them to potential adversaries. }
\subsubsection{Integration of Differential Privacy}
\rehmani{The notion of differential privacy has been well researched in literature from the perspective of IIoT~\cite{crnapp16}. These research works have proved that integration of differential privacy with IIoT sensors and nodes can result in a fruitful outcome. For example, during sharing of mutual information among sensors, differential privacy perturbation can protect sensors and involve user’s privacy. Similarly, these concepts can also be applied to CR-based IIoT networks because they share pretty much similar space. For example, if an IIoT node has to carry out SS to choose the most viable spectrum band, then it can perturb its identity and location values to protect itself from network adversaries. Similarly, during information sharing on unlicensed spectrum, these IIoT nodes can add noise to protect information of users being shared via their medium. That is why, we believe that adding differential privacy with a CR-based IIoT network will pave the way for futuristic IIoT systems.}

\subsection{Blockchain based Cognitive Radio Networks}
\rehmani{Blockchain came into sight as a backbone technology after the sudden boom in the price of Bitcoin since the past decade~\cite{crnapp17}. Since then, research works are being carried out to integrate blockchain in almost every second aspect of our everyday life~\cite{crnapp18}. In the quest of this integration, blockchain technology is now being used to perform various CR operations. For example, the steps involved in the cognition cycle are being performed on decentralized distributed blockchain networks in order to enhance trust in the network. This recording of CRN data on blockchain ledger ensures that all participants receive a fair outcome and will prevail a sense of security and trust in the network~\cite{mubcrn02}. However, this decentralized nature also raises various questions, and one of the biggest questions among them is privacy of CR nodes. As mentioned, that the data from CR nodes will be reported and recorded on a tamper-proof decentralized ledger, which means that this data will be visible to all participating nodes in case of a public blockchain to ensure trust, but on the other hand an adversary or a node with adversarial intentions can also misuse this data as well. Similarly, as the data is tamper-proof and will always be there on the chain, then some adversarial node can also learn about past history of a CR node even if the node has left the network. These are certain privacy aspects which need to be addressed in detail before practical deployment of blockchain with CRN.}
\subsubsection{Integration of Differential Privacy}
\rehmani{To preserve privacy of blockchain based CRN, differential privacy can be a feasible solution because of the diverse adaptability of differential privacy protection. The biggest reason that makes blockchain based CRN prone to privacy attack is the public availability of plain-text data on decentralized distributed ledger. This issue can be countered by adding pseudo random noise via differential privacy to the data before reporting/recording it to a decentralized ledger. The noise in differential privacy is controlled by privacy budget, which ensures that the noisy values remain useful to carry out various operations such as statistical analysis, etc. But on the other hand, the strong theoretical guarantee of differential privacy also ensures that the adversary should not be able to get private information of CR nodes from the recorded data. Similarly, in private blockchain networks, certain analyses have been carried out to learn from CR data for futuristic purposes. \cogcom{This analysis can also be secured by introducing a layer of differential privacy between blockchain ledger and observer. Considering this discussion, we believe that differential privacy is one of the critical mechanisms which can play a key role in development of modern blockchain based CRN.}}

\subsection{Cognitive Radio Based Vehicular Networks}
\rehmani{\cogcom{Vehicular communication is not a new topic and it has been in the scientific community since ages~\cite{crnapp19}. Similar to other networks, these vehicular networks have also been made capable of running over unlicensed CR spectrum in order to preserve excessive usage of spectrum~\cite{crnapp20}.} A plethora of research has been carried out to perform all operations of vehicular networks via CR so that modern vehicles do not contribute to spectrum shortage~\cite{crnapp21}. These research works have indicated that this integration of CR with vehicular ad hoc network is a viable solution, and vehicles can achieve low-latency rate and ultra-reliable communication by incorporating CRN during their communication. Although, on the other side of the coin, these CR-based vehicular ad hoc networks (VANETS) are not completely secure and are prone to various privacy attacks because of their real-time reporting. Therefore, protecting privacy of these CR-based VANETS is of sheer importance and this issue needs to be resolved. }
\subsubsection{Integration of Differential Privacy}
\rehmani{To protect privacy of CR-based VANETS, it is important to analyse the type of privacy leakages among these networks. If we analyse, it is evident that the most critical private information among CR-based VANETS is location privacy of vehicles. For example, one does not want to show to others whether he/she visited the hospital at a particular time or not. However, in case of CR-based VANET reporting, sensing, and analysis, this information is prone to adversaries, and any adversary with some background knowledge can get insights of exact values. In order to overcome this, differential privacy can be used to protect location privacy. As discussed in Section~\ref{Sec:Integration}, one can protect location information pretty easily by just incorporating randomness via differential privacy models. For example, if a vehicle adds differentially private noise to its location coordinated during SS, then the original coordinates will become protected. However, on the other hand the SS results will not have much effect because of the vehicle being present in the same region. Therefore, we believe integration of differential privacy with CR-based VANETS can serve as a viable step towards development of more secure and private vehicular networks. }









\section{Challenges and Future Research Directions} \label{Sec:Challenge}
\mubashir{Till now, we provide a detailed overview of how differential privacy can play a critical role in various aspects of the CRN cognition cycle. For instance, we highlight the integration scenarios of differential privacy in SS, analysis, sharing, and mobility. Similarly, we highlight various parameters that need to be taken care of while designing differentially private CRN models. However, apart from all these discussions, there are certain challenges and future directions that need special considerations while designing future differentially private CRN models. In this section, we provide first provide insights about certain prospective integrations that can be beneficial for differentially private CRN, and then we highlight certain challenges that researchers exploring these directions can face during their evaluations. }

\subsection{Integrating Blockchain with Differential Privacy and CRN}
\mubashir{During cognition cycle of CRN, SU and PU nodes suffer with lack of trust due to centralized entities. For instance, during database-drive SS, one has to rely on a centralized database to provide efficient results. Similarly, this centralization is also a part of other steps in the cognition cycle, and it cannot be ignored. Indeed, this centralization provides benefits such as quick response, etc, but on the other hand it also raises serious trust-related issues in the network. Therefore, there is a need to develop decentralized models for modern CRN~\cite{newchal01}. In order to provide an efficient alternative to centralization, the notion of blockchain came up as a saviour for CRN.  Blockchain is a novel paradigm which is being applied to a large number of daily life domains because of its trust, availability, and tamper-proof nature~\cite{crnchallenge01}. Similarly, certain works also discussed and evaluated the integration of blockchain with CRN ranging from SS to other steps of the cognition cycle~\cite{crnchallenge02, crnchallenge03, crnchallenge04, crnchallenge05, crnchallenge06}. However, just integrating blockchain with CRN is not a one-in-all solution to every problem as it also suffers from privacy issues due to its public nature~\cite{challenge07}. Therefore, it is important to protect privacy of decentralized blockchain CRN.}\\
\mubashir{In order to ensure privacy in decentralized blockchain based CRN, we believe that differential privacy can play a critical role. A very detailed survey on integration of differential privacy in various layers of blockchain has been presented in~\cite{crnchallenge08}. \cogcom{However, the aspect of integration of differential privacy in blockchain based CRN has not been carried out yet. From the works integrating differential privacy with blockchain, it can be seen that it is an efficient solution to the privacy problem, but it still requires detailed research. For instance, one of the critical challenges for this integration is to choose optimal blockchain} type, e.g., there could be scenarios in which public blockchain will perform the best as compared to consortium or private. Contrarily, for some functionalities private or consortium blockchain will outperform public blockchain. Similarly, choosing an optimal differential privacy budget according to the decentralized nature of blockchain is another challenge that needs to be discussed in future works. Overall, we believe that this integration of blockchain, differential privacy, and CRN have a lot of scope in future but certain aspects require in-depth addressing before practical implementation.}

\subsection{Differential Privacy in Game-Theoretic Spectrum Sharing Models}
Since, spectrum is a non-renewable resource, therefore, it is important to use and allocate the spectrum in the most efficient manner. \cogcom{In order to do so, certain researchers worked over integration of game-theory in various scenarios of the cognition cycle. For instance, some works highlighted use dynamic games to get maximum benefits from CRN~\cite{crnchallenge09}. Similarly, certain works highlighted use repeated games for power/spectrum allocation in CRN to maximize spectrum usage efficiency~\cite{crnchallenge10}.} Alongside this, certain works also highlighted to use game-theory based auctions for CR spectrum trading~\cite{crnsource33}. However, as the recent studies showed that CRN is vulnerable to certain privacy issues, therefore, it is important to integrate privacy preservation mechanisms with game-theoretic CRN approaches. The dynamic functionality of differential privacy can play a vital role in this integration, e.g., one aspect could be to design game-theoretic differentially private auctions that maximizes revenue alongside preserving privacy. This aspect has been touched by certain research works that we discussed in previous sections. \mubashir{However, it is important to highlight that maximizing revenue while preserving privacy is one of the biggest challenges that these mechanisms face. For instance, in a simple game-theoretic auction revenue can be maximized by proving various equilibriums in the system, such as Nash equilibrium, etc. However, in differentially private auctions, we cannot publicize the bids, which becomes a big hurdle to find optimal values. Certain works focused over approximate social welfare/revenue maximization, but this field still has a lot of potential which needs to be explored further.}\\
\mubashir{Another direction could be to integrate differential privacy with resource allocation of CRN, in which efficient resource allocation/sharing can be carried out in a private manner. This aspect of traditional resource allocation while maintaining equilibrium has been discussed by researchers~\cite{newchal02}. However, from the perspective of CRN, this problem is not well addressed, and we need modern differentially private strategies specifically designed for game-theoretic resource allocation by considering dynamic spectrum access capability of CRN. A significant challenge that one can face while designing these differentially private CRN allocation strategies is to perform truthful reporting for any game-theoretic mechanism. Therefore, researchers should focus over this direction in order to get maximum benefit from game-theory for differentially private CRN.}

\subsection{\cogcom{Differentially Private Cognitive Radio Trade-offs}}
\cogcom{Since, CR works over the phenomenon of utilization of unused spectrum band when PU is not using the specific band. Thus, in order to utilize the spectrum in the most efficient manner, SUs are continuously sensing the environment to get better opportunities, and they switch to the best available spectrum. However, during this cognition cycle process, a large number of processes are taking place, which sometime can lead to unwilling circumstances, such as false alarm, energy wastage, etc. Thus, in order to eradicate such a phenomenon, SUs try to opt certain strategies to ensure the effectiveness of sensing, e.g., increasing sensing time duration for a specific band/area, etc. These strategies help in efficient detection, but there are certain trade-offs associated with this. E.g., one of the most famous trade-offs is sensing-throughput trade-off, in which detection probability and false alarm probability is taken into account in order to figure out the efficiency of the specific CRN~\cite{cogcomref10}. Another famous trade-off for CRN is energy efficiency and spectral efficiency trade-off, where different architectures (such as cooperative, non-cooperative, etc.), links, analyses, and probabilities are considered to figure out efficient balance point among energy and spectral efficiency~\cite{cogcomref11}.}\\
\cogcom{From the perspective of relevance of these trade-offs with privacy preservation, we believe that it is important to highlight that majority of these trade-offs can be linked to leakage of privacy in certain aspects. E.g., in sensing-throughput trade-off, in order to enhance the throughput, sensing time needs to be increased, which in turn leads into learning a lot more than required about a specific spectrum/coverage area, which directly leads to various privacy leakages. Similarly, in energy related trade-offs, selecting a specific architecture or relay model, etc. also require learning about a specific area, which leads to leakage of privacy. Therefore, it is important to also consider the prospect of privacy preservation while observing these trade-offs. In order to protect privacy during these trade-offs, we believe that differential privacy can play an active role. For example, the location and identity privacy during excessive SS can easily be preserved via Laplacian and Exponential perturbation, in which we perturb the values to ensure that individuals cannot be re-identified, but on the other hand efficiency and accuracy is also taken into consideration in order to enhance throughput. Similarly, differential privacy can also perturb the learned values for a specific architecture in such a way that privacy is maintained alongside ensuring the accuracy and effectiveness. }

\subsection{Differential Privacy in Spectrum Characterization }
Spectrum characterization is the core foundation of the spectrum analysis step during the cognitive cycle. In spectrum characterization, all available spectrum bands and channels are collected and categorized according to the need~\cite{crnchallenge11}. This step ensures that characteristics of all available spectrum bands are gathered and grouped in an efficient way \maaz{to get maximum benefits from available spectrum. In order to achieve this, researchers} are also integrating modern machine/deep learning mechanisms to group these values. However, this advancement can also cause privacy leakage during the learning process. As it can be seen from experimental evaluations that machine learning algorithms can be adversely used to learn about characteristics of participants~\cite{crnchallenge12}. Same is the case with CRN, that if we integrate machine/deep learning during spectrum characterization, then privacy can be leaked in an adversarial manner.\\
In order to mitigate this privacy risk, differential privacy can play an active role due to its dynamic nature. For instance, one can develop a differentially private machine learning model to train and group spectrum bands in an efficient manner. This integration of differential privacy ensures that the added noise is pseudo random, and no one can extract private information of participants by just looking into it. \mubashir{However, this integration is not as simple as it seems because it involves a large number of challenges that needs to be tackled. First of all, one has to choose an optimal machine/deep learning model that matches perfectly with the nature of dynamic differential privacy and CRN. Afterwards, the second big challenge is to figure out parameter training values via which we get both utility and privacy during characterization. Keeping in view this discussion, it can be said that differentially private spectrum characterization can provide us a lot of benefits, but there are certain challenges that need to be addressed before this fruitful integration.}

\subsection{Differential Privacy in CRN for Smart Grid System}
CRN has been integrated with smart grid technology for a long time due to its numerous benefits such as low communication cost~\cite{crnchallenge13}. If one analyses this integration, it can be seen that CRN is being used in almost all aspects of \maaz{smart grid. For instance,} certain works analysed integration of CRN in home management, while other works discussed collection of meter reading, pricing, and power outage values via CRN. Similarly, wide area monitoring and power line monitoring is also being carried out through CR-based smart grid. A very detailed survey on the integration of CR with smart grid has been written by Athar~\textit{et al.}~\cite{crnchallenge14}. From these integrations it is evident that CR is paving a way for futuristic smart grids. But on the other hand, the privacy issues during this integration cannot be ignored. \\
Ranging from privacy leakage from real-time energy monitoring to dynamic energy auctions, the communication is vulnerable, and it needs to be protected from adversaries. In order to do so, differential privacy can play a key role because of its randomization mechanisms. Similar to this, differential privacy has been applied with smart grid scenarios since long, such as differentially private smart metering~\cite{crnchallenge15}. But the works integrating differential privacy with CRN-based smart grid are not yet covered. \mubashir{Therefore, there is a need to carry out this integration of differential privacy with CRN based smart grid in order to get maximum possible benefits from these advancing technologies. Nevertheless, this integration is pretty beneficial, but this will raise certain challenges as well, the most important among them will be maintaining the balance between utility and privacy in the protected data. The values reported by smart grid systems involve decisions related to energy, therefore, it is important to take special care of data utility because even a very small mistake can lead to catastrophic events. Therefore, the researchers who are intended to work in differentially private CRN based smart grids have to deal with the challenge of finding optimal privacy budget, sensitivity, and other parametric values for differential privacy mechanisms. }

\subsection{Integrating Differential Privacy in Federated Learning with CRN}
Federated learning is also a novel paradigm which is being used to carry out decentralized learning at users' end without collecting data from them~\cite{crnchallenge16}. \maaz{Federated learning is being applied to various scenarios of cognitive cycle~\cite{crnchallenge18}.} This concept of federated learning is providing the advantage of decentralized learning in CRN, and now FCs can learn about the characteristics and availability of spectrum without collecting sensitive information from CR nodes. This enhances the security and privacy of the whole CRN because minimal data is being collected at a centralized server.\\
Research works have shown that this integration can also cause privacy leakage, and even in some worst case scenarios adversarial federated learning models can be used by adversaries to learn private information from users~\cite{crnchallenge19, crnchallenge20}. In order to overcome this risk, the integration of differential privacy with federated learning models can play a key role. For instance, certain works have been developed which have integrated differential privacy with federated learning at the time of model design, and learning. Similarly, a differentially private federated learning model can be designed for the learning environment of CRN, or differential privacy can be integrated with learning outcomes during the run-time. Both of these scenarios can be considered to preserve the privacy of CRN. \mubashir{Although, both of these scenarios can be pretty beneficial to preserve privacy, but one also has to overcome certain challenges while considering these scenarios. The most  important challenge is to figure out the best type of federated learning model. For instance, two most prominent models for federated learning are horizontal federated learning and federated transfer learning. Both of these models have their pros and cons, e.g., if the datasets being used in federated learning share the same feature then horizontal federated learning is the optimal way. Contrarily, if there are not many overlapping features among data, then federated transfer learning is used. Thus, in differentially private CRN, for some scenarios, such as SS, horizontal federated learning is more suitable, because the shared features are the same, but in some cases this approach might not be feasible. \cogcom{E.g., in case of spectrum hand-off, the responses after spectrum mobility can comprise multiple disjoint features. So, in such cases federated transfer learning is more suitable. Therefore, finding optimal differentially private federated learning models is the key challenge that needs to be resolved before this integration.}}

\section{Conclusion} \label{Sec:Conclusion}
\maaz{Spectrum is a non-renewable resource. It is therefore important to use this precious resource in an efficient manner. In order to carry out efficient utilization of spectrum, scientists developed the notion of CR, which works over the principle of spectrum access at vacant times. CR nodes have the ability to sense the loopholes in the spectrum and then use these loopholes to carry out communication. In this way, CR nodes can play a vital role in overcoming spectrum scarcity. Nevertheless, CRN has a large number of benefits, they are not immune to all threats and one of the most critical ones is the privacy leakage, which causes serious consequences} if not handled properly. Certain research works have highlighted the use of various privacy preservation approaches to protect privacy of CRN, and differential privacy is one of them. \maaz{Differential privacy can play an important role in the design and development of modern, private, and more secure CRN of the future. In this paper, we carried out a comprehensive survey targeting the integration of differential privacy in CRN from various aspects. Firstly, we highlight the importance of privacy preservation in CRN, by discussing the functioning of differential privacy. We then provide an in-depth discussion about the sources of privacy leakage in CRN. Next} we provide insights into how differential privacy can play a critical role in protecting this leakage. \maaz{We then present an analysis of certain parameters that should be taken} into account while developing differential privacy based CRN protocols. Then, an in-depth analysis of technical works integrating differential privacy in various scenarios of CRN. \maaz{Finally, we provide analysis about prospective future directions alongside highlighting certain challenges that researchers may face.}

\section{Compliance with Ethical Standards}

\subsection{Funding}
No funding sources are associated with this article.

\subsection{Conflict of Interest}
Author Muneeb Ul Hassan declares that he has no conflict of interest. Author Mubashir Husain Rehmani declares that he has no conflict of interest. Author Maaz Rehan declares that he has no conflict of interest. Author Jinjun Chen declares that he has no conflict of interest.

\subsection{Ethical Approval}
This article does not contain any studies with human participants or animals performed by any of the authors.

\bibliographystyle{IEEEtran}


\end{document}